\documentclass[preprint,prd,nofootinbib,showpacs]{revtex4}
\usepackage{graphicx}
\usepackage{epsfig}
\usepackage{bm}
\begin{document}   

\title{\large\bf Model independent analysis of $b \to (c,\,u)\,\tau\nu$ leptonic and semileptonic decays}
\author{Rupak~Dutta${}^{1}$}
\email{rupak@phy.nits.ac.in}
\author{Anupama~Bhol${}^{2}$}
\email{anupama.phy@gmail.com}   
\affiliation{
${}^1$National Institute of Technology Silchar, Silchar 788010, India\\
${}^2$C.~V.~Raman College of Engineering, Bhubaneswar, Odisha 752054, India 
}

\begin{abstract}
Latest measurement of the ratio of branching ratios $R_D =  \mathcal B(B \to D\,\tau\,\nu)/\mathcal B(B \to D\,l\,\nu)$ 
and $R_{D^{\ast}} =  \mathcal B(B \to D^{\ast}\,\tau\,\nu)/\mathcal B(B \to D^{\ast}\,l\,\nu)$, where $l$ is either an electron 
or muon, differs from the standard model expectation by $1.9\sigma$ and $3.3\sigma$, respectively. Similar tension has been 
observed in purely leptonic $B \to \tau\nu$ decays as well. In this context, we consider the most general effective Lagrangian 
in the presence of new physics and perform a model independent analysis to explore various new physics couplings. 
Motivated by the recently proposed new observables $R_D^{\tau} = R_D/\mathcal B(B \to \tau\nu)$ and $R_{D^{\ast}}^{\tau} = 
R_{D^{\ast}}/\mathcal B(B \to \tau\nu)$, we impose $2\sigma$ constraints coming from $R_D^{\tau}$ and $R_{D^{\ast}}^{\tau}$ 
in addition to the constraints coming from $R_D$, $R_{D^{\ast}}$, and $\mathcal B(B \to \tau\nu)$ to constrain the new physics 
parameter space. We study the impact of new physics on various observables related to $B_s \to (D_s,\,D^{\ast}_s)\tau\nu$ and 
$B \to \pi\tau\nu$ decay processes.
\end{abstract}


\maketitle

\section{Introduction}
\label{int}
Although the anomalies in the $B$ meson decays suggest presence of new physics~(NP) in the
flavor sector, NP is yet to be confirmed. Various model dependent as well as model 
independent analysis have been carried out to explore different NP scenarios. More 
specifically, the $b \to u$ and $b \to c$ leptonic and semileptonic decays of $B$ meson 
such as $B \to (D,\, D^{\ast})\,\tau\nu$, $B \to \pi\tau\nu$, and $B \to \tau\nu$ decays have been the center 
of attraction among the physics communities in the last few years~\cite{Fajfer1, Fajfer2, Hou, Akeroyd, tanaka, 
Nierste, miki, Wahab, Deschamps, Blankenburg, Ambrosio, Buras, Pich, Jung, Crivellin, datta, datta1, datta2, datta3, 
datta4, fazio, Crivellin1,  Celis, He, dutta, Tanaka:2016ijq, Deshpand:2016cpw,  Li:2016vvp, Du:2015tda, 
Bernlochner:2015mya, Soffer:2014kxa, Bordone:2016tex, Bardhan:2016uhr, Alok:2016qyh, Ivanov:2015tru, Ivanov:2016qtw, 
Boucenna:2016wpr, Boucenna:2016qad}. Of late, various baryonic decay modes such as 
$\Lambda_b \to \Lambda_c\,\tau\,\nu$ and $\Lambda_b \to p\,\tau\,\nu$ mediated via $b \to (c,\,u)\tau\nu$ transition 
processes also got some attention because of the high production of $\Lambda_b$ at the 
LHC~\cite{Woloshyn:2014hka, Shivashankara:2015cta, Gutsche:2015mxa, Detmold:2015aaa, Dutta:2015ueb}. The 
semileptonic $B$ decays are sensitive probes to search for various NP models such as two 
Higgs doublet model~(2HDM), minimal suppersymmetric standard model~(MSSM) and leptoquark 
model.  Exclusive semileptonic $B$ decays was first observed by BELLE collaboration~\cite{Matyja:2007kt}, with subsequent 
studies reported by BELLE~\cite{Bozek:2010xy, Huschle:2015rga} and BABAR~\cite{Lees:2012xj, Lees:2013uzd}. The recent measurement on the 
ratio of branching ratios $R_D$ and $R_{D^{\ast}}$ are
\begin{eqnarray}
&&R_D^{\rm BABAR} = 0.440\pm 0.058\pm 0.042\,, \qquad\qquad
R_{D^{\ast}}^{\rm BABAR} = 0.332\pm 0.024\pm 0.018\,, \nonumber \\
&&R_D^{\rm BELLE} = 0.375\pm 0.064\pm 0.026\,, \qquad\qquad
R_{D^{\ast}}^{\rm BELLE} = 0.293\pm 0.038\pm 0.015\,,
\end{eqnarray}
where the first uncertainty is statistical and the second one is systematic. Very recently
LHCb has also measured the ratio $R_{D^{\ast}}$ to be $0.336\pm 0.027\pm 0.030$~\cite{Aaij:2015yra}.
Again, BELLE has reported their latest measurement on $R_{D^{\ast}} = 0.302 \pm 0.030 \pm 0.011$ with a semileptonic 
tagging method~\cite{Sato:2016svk} which is within $1.6\sigma$ of the standard model~(SM) theoretical expectation.
The measured values of $R_D$ and $R_{D^{\ast}}$ exceed the SM prediction by
$1.9\sigma$ and $3.3\sigma$ respectively. Considering the $R_D$ and $R_{D^{\ast}}$ 
correlation, the combined analysis of $R_D$ and $R_{D^{\ast}}$ finds the deviation from the 
SM prediction to be at more than $4.0\sigma$ level~\cite{Amhis:2014hma}. The combined results from the leptonic 
and hadronic decays of $\tau$, the BABAR and BELLE measured value of $\mathcal B(B \to \tau\nu)$ are 
$(1.83^{+0.53}_{-0.49})\times 10^{-4}$~\cite{Lees:2012ju} and $(1.25\pm0.28)\times 10^{-4}$~\cite{Kronenbitter:2015kls}, 
respectively. BELLE measurement is consistent with the SM prediction for both exclusive and inclusive $V_{ub}$, whereas, 
with the exclusive $V_{ub}$, there is still some discrepancy between the BABAR measured value of 
$\mathcal B(B \to \tau\nu)$ and the SM theoretical prediction.

Very recently, in Ref.~\cite{Nandi:2016wlp}, various new observables such as $R^{\tau}_D$ and 
$R^{\tau}_{D^{\ast}}$ have been proposed to explore the correlation between the new 
physics signals in $B \to (D,\,D^{\ast})\tau\nu$ and $B \to \tau\nu$ decays. These observables 
\begin{eqnarray}
&&R^{\tau}_{D} = \frac{R_D}{\mathcal B(B \to \tau\nu)}\,, \qquad\qquad
R^{\tau}_{D^{\ast}} = \frac{R_{D^{\ast}}}{\mathcal B(B \to \tau\nu)}
\end{eqnarray} 
are obtained by dividing the ratio of branching ratios $R_D$ and $R_{D^{\ast}}$ by $B \to \tau\nu$ branching ratio. 
Although, $\tau$ detection and identification systematics are present in $B \to \tau\nu$ and $B \to (D,\,D^{\ast})\tau\nu$
decays, it will mostly cancel in these newly constructed ratios. However, these ratios suffer from large 
uncertainties due to the 
presence of not very well known parameter $V_{ub}$ in the denominator.
The estimated values are~\cite{Nandi:2016wlp}
\begin{eqnarray}
&&R_D^{\tau^{\rm BABAR}}(\times {10}^3) = 2.404\pm 0.838\,, \qquad\qquad
R_{D^{\ast}}^{\tau^{\rm BABAR}}(\times {10}^3) = 1.814\pm 0.582\,, \nonumber \\
&&R_D^{\tau^{\rm BELLE}}(\times {10}^3) = 3.0\pm 1.1\,, \qquad\qquad
R_{D^{\ast}}^{\tau^{\rm BELLE}}(\times {10}^3) = 2.344\pm 0.799\,.
\end{eqnarray}
The estimated values of these new observables from BABAR and BELLE measured values of the ratio of branching ratios
$R_D$, $R_{D^{\ast}}$, and $\mathcal B(B \to \tau\nu)$ are consistent with the SM 
prediction~\cite{Nandi:2016wlp} although the measured values of $R_D$ and $R_{D^{\ast}}$ itself differ 
from the SM prediction. It, however, does not necessarily rule out the possibility of 
presence of NP because even if NP is present, the effect of it may largely cancel in the 
ratios. In Ref.~\cite{Nandi:2016wlp}, the authors discuss the constraints on 2HDM parameter space 
using the constraints coming from the estimated values of $R^{\tau}_D$ and 
$R^{\tau}_{D^{\ast}}$ and find that although the BABAR data does not allow a simultaneous 
explanation of all the above mentioned deviations, however, for BELLE data, there actually
a common allowed parameter space. In this present study, we use the most general effective Lagrangian in the presence
of NP to study various NP effects on $b \to u$ and $b \to c$ 
leptonic and semileptonic decays. First, we consider the constraints coming from the 
measured values of $R_D$, $R_{D^{\ast}}$, and $\mathcal B(B \to \tau\nu)$ to explore 
various NP effect. Second, we see whether it is 
possible to constrain the NP parameter space even further by putting additional 
constraints coming from the estimated values of $R^{\tau}_D$ and $R^{\tau}_{D^{\ast}}$ 
since the estimated values of these ratios are consistent with the SM values. We also give prediction 
on other similar observables related to $B \to \pi\tau\nu$ and $B_s \to (D_s,\,D^{\ast}_s)\,\tau\nu$ decays.

In section~\ref{ehha}, we start with a brief description of the effective Lagrangian for 
the $b \to (u,\,c)\,l\,\nu$ transition decays in the presence of NP. All the relevant formulas such as the partial decay width of 
$B \to l\,\nu$ decays and differential decay width of three body $B \to (P,\,V)\,l\,\nu$ decays 
are reported in section~\ref{ehha}. We also construct various new observables related to semileptonic $B$ and 
$B_s$ meson decays. In section~\ref{rd}, we start with the input parameters that are used for our numerical computation. 
The SM prediction and the effect of each NP couplings on various observables related to semileptonic $B$ and $B_s$ meson decays are  
reported in section~\ref{rd}. We conclude with a brief summary of our results in 
section~\ref{con}.
 
\section{Helicity amplitudes within effective field theory approach}
\label{ehha}
In the presence of NP, the effective weak Lagrangian for the $b \to q^{\prime}\,l\,\nu$ 
transition decays, where $q^{\prime}$ is either a $u$ quark or a $c$ quark, can be written
as~\cite{Bhattacharya, Cirigliano}
\begin{eqnarray}
\mathcal L_{\rm eff} &=&
-\frac{4\,G_F}{\sqrt{2}}\,V_{q^\prime b}\,\Bigg\{(1 + V_L)\,\bar{l}_L\,\gamma_{\mu}\,\nu_L\,\bar{q^\prime}_L\,\gamma^{\mu}\,b_L +
V_R\,\bar{l}_L\,\gamma_{\mu}\,\nu_L\,\bar{q^\prime}_R\,\gamma^{\mu}\,b_R 
+
\widetilde{V}_L\,\bar{l}_R\,\gamma_{\mu}\,\nu_R\,\bar{q^\prime}_L\,\gamma^{\mu}\,b_L \nonumber \\
&&+
\widetilde{V}_R\,\bar{l}_R\,\gamma_{\mu}\,\nu_R\,\bar{q^\prime}_R\,\gamma^{\mu}\,b_R +
S_L\,\bar{l}_R\,\nu_L\,\bar{q^\prime}_R\,b_L +
S_R\,\bar{l}_R\,\nu_L\,\bar{q^\prime}_L\,b_R +
\widetilde{S}_L\,\bar{l}_L\,\nu_R\,\bar{q^\prime}_R\,b_L +
\widetilde{S}_R\,\bar{l}_L\,\nu_R\,\bar{q^\prime}_L\,b_R \nonumber \\
&&+ 
T_L\,\bar{l}_R\,\sigma_{\mu\nu}\,\nu_L\,\bar{q^\prime}_R\,\sigma^{\mu\nu}\,b_L +
\widetilde{T}_L\,\bar{l}_L\,\sigma_{\mu\nu}\,\nu_R\,\bar{q^\prime}_L\,\sigma^{\mu\nu}\,b_R\Bigg\} + {\rm h.c.}\,,
\end{eqnarray}
where, $G_F$ is the Fermi coupling constant and $V_{q^{\prime}b}$ is the CKM matrix element. The vector, scalar, and tensor type 
NP interactions denoted by $V_{L,\,R}$, $S_{L,\,R}$, and $T_L$ are associated with left handed neutrinos, whereas, 
$\widetilde{V}_{L,\,R}$, $\widetilde{S}_{L,\,R}$, and $\widetilde{T}_L$ type NP couplings are associated with right 
handed neutrinos. We consider all the NP couplings to be real for our analysis. Again, we keep only vector and scalar 
type NP couplings in our analysis. We rewrite the effective Lagrangian as~\cite{dutta}
\begin{eqnarray}
\label{leff}
\mathcal L_{\rm eff} &=&
-\frac{G_F}{\sqrt{2}}\,V_{q^\prime b}\,\Bigg\{G_V\,\bar{l}\,\gamma_{\mu}\,(1 - \gamma_5)\,\nu_l\,\bar{q^\prime}\,\gamma^{\mu}\,b -
G_A\,\bar{l}\,\gamma_{\mu}\,(1 - \gamma_5)\,\nu_l\,\bar{q^\prime}\,\gamma^{\mu}\,\gamma_5\,b +
G_S\,\bar{l}\,(1 - \gamma_5)\,\nu_l\,\bar{q^\prime}\,b \nonumber \\
&& - G_P\,\bar{l}\,(1 - \gamma_5)\,\nu_l\,\bar{q^\prime}\,\gamma_5\,b + 
\widetilde{G}_V\,\bar{l}\,\gamma_{\mu}\,(1 + \gamma_5)\,\nu_l\,\bar{q^\prime}\,\gamma^{\mu}\,b -
\widetilde{G}_A\,\bar{l}\,\gamma_{\mu}\,(1 + \gamma_5)\,\nu_l\,\bar{q^\prime}\,\gamma^{\mu}\,\gamma_5\,b \nonumber \\
&&+
\widetilde{G}_S\,\bar{l}\,(1 + \gamma_5)\,\nu_l\,\bar{q^\prime}\,b - \widetilde{G}_P\,\bar{l}\,(1 + \gamma_5)\,\nu_l\,\bar{q^\prime}\,
\gamma_5\,b\Bigg\} + {\rm h.c.}\,,
\end{eqnarray}
where 
\begin{eqnarray*} 
&&G_V = 1 + V_L + V_R\,,\qquad\qquad
G_A = 1 + V_L - V_R\,, \qquad\qquad
G_S = S_L + S_R\,,\qquad\qquad
G_P = S_L - S_R\, \nonumber \\
&&\widetilde{G}_V = \widetilde{V}_L + \widetilde{V}_R\,,\qquad\qquad
\widetilde{G}_A = \widetilde{V}_L - \widetilde{V}_R\,, \qquad\qquad
\widetilde{G}_S = \widetilde{S}_L + \widetilde{S}_R\,,\qquad\qquad
\widetilde{G}_P = \widetilde{S}_L - \widetilde{S}_R\,.
\end{eqnarray*}
The SM contribution can be obtained once we set $V_{L,R} = S_{L,R} = \widetilde{V}_{L,R} =
\widetilde{S}_{L,R}=0$ in Eq.~(\ref{leff}). In the presence of NP, the partial decay width
of $B \to l\,\nu$ and differential decay width of three body $B_{q} \to (P,\,V)\,l\,\nu$ decays, where $P$ is a pseudoscalar meson 
and $V$ is a vector meson can 
be expressed as~\cite{dutta}
\begin{eqnarray}
\Gamma(B \to l\nu) &=&
\frac{G_F^2\,|V_{ub}|^2}{8\,\pi}\,f_B^2\,m_l^2\,m_{B}\,\Big(1 - \frac{m_l^2}{m_B^2}\Big)^2\,
\Bigg\{\Big[G_A - \frac{m_B^2}{m_l\,(m_b(\mu) + m_u(\mu))}\,G_P\Big]^2 \nonumber \\
&&+ 
\Big[\widetilde{G}_A - \frac{m_B^2}{m_l\,(m_b(\mu) + m_u(\mu))}\,\widetilde{G}_P\Big]^2\Bigg\}\,,
\end{eqnarray}
\begin{eqnarray}
\label{pilnu}
\frac{d\Gamma^P}{dq^2} &=&
\frac{8\,N\,|\overrightarrow{p}_P|\,}{3}\Bigg\{\,H_0^2\,\Big(G_V^2 + \widetilde{G}_V^2\Big)\,\Big(1 + \frac{\,m_l^2}{2\,q^2}\Big) \nonumber\\
&& + \frac{3\,m_l^2}{2\,q^2}\,\Big[ \Big(H_t\,G_V + \frac{\sqrt{q^2}}{m_l}\,H_S\,G_S \Big)^2 + \Big(H_t\,\widetilde{G}_V + \frac{\sqrt{q^2}}
{m_l}\,H_S\,\widetilde{G}_S\Big)^2\Big] \Bigg\}\,
\end{eqnarray}
and
\begin{eqnarray}
\label{vlnu1}
\frac{d\Gamma^V}{dq^2} &=&
\frac{8\,N\,|\overrightarrow{p}_V|}{3}\,\Bigg\{ \mathcal{A}_{AV}^2 + \, \frac{ m_l^2}{2\,q^2}\Big[ \mathcal{A}_{AV}^2 + 3\mathcal{A}_{tP}^2 
\Big] 
+
\widetilde{\mathcal{A}}_{AV}^2 + \, \frac{m_l^2}{2\,q^2}\Big[ \widetilde{\mathcal{A}}_{AV}^2 + 3\mathcal{\widetilde{A}}_{tP}^2 \Big] \Bigg\}
\end{eqnarray}
where 
\begin{eqnarray}
&& |\overrightarrow{p}_{(P,\,V)}| = \sqrt{\lambda(m_{B_q}^2,\,m_{(P,\,V)}^2,\,q^2)}/2\,m_{B_q}\,, \qquad\qquad
\lambda(a,\,b,\,c) = a^2 + b^2 + c^2 - 2\,(a\,b + b\,c + c\,a) \nonumber\\
&&N = \frac{G_F^2\,|V_{q^\prime b}|^2\,q^2}{256\,\pi^3\,m_{B_q}^2}\,\Big(1 - \frac{m_l^2}{q^2}\Big)^2\,, \qquad\qquad
H_0 = \frac{2\,m_{B_q}\,|\overrightarrow{p}_P|}{\sqrt{q^2}}\,F_{+}(q^2) \nonumber \\
&&H_t = \frac{m_{B_q}^2 - m_P^2}{\sqrt{q^2}}\,F_0(q^2)\,, \qquad\qquad
H_S=\frac{m_{B_q}^2 - m_P^2}{m_b(\mu) - m_{q^\prime}(\mu)}\,F_0(q^2)\,,\nonumber \\
&&\mathcal{A}_{AV}^2 = \mathcal{A}_0^2\,G_A^2 + \mathcal{A}_\parallel^2\,G_A^2 + \mathcal{A}_\perp^2\,G_V^2 \,, \qquad\qquad
\widetilde{\mathcal{A}}_{AV}^2=\mathcal{A}_0^2\,\widetilde{G}_A^2 + \mathcal{A}_\parallel^2\,\widetilde{G}_A^2 + \mathcal{A}_\perp^2\,
\widetilde{G}_V^2 \nonumber\\
&&\mathcal{A}_{tP}=\mathcal{A}_t\,G_A + \frac{\sqrt{q^2}}{m_l}\,\mathcal{A}_P\,G_P \,,\qquad\qquad
\mathcal{\widetilde{A}}_{tP}=\mathcal{A}_t\,\widetilde{G}_A + \frac{\sqrt{q^2}}{m_l}\,\mathcal{A}_P\,\widetilde{G}_P \,.
\end{eqnarray}
and
\begin{eqnarray}
&&\mathcal{A}_0=\frac{1}{2\,m_V\,\sqrt{q^2}}\Big[\Big(\,m_{B_q}^2-m_V^2-q^2\Big)(m_{B_q}+m_V)A_1(q^2)\,-\,
\frac{4m_{B_q}^2|\vec p_V|^2}{m_{B_q}+m_V}A_2(q^2)
\Big]\,, \nonumber\\
&&\mathcal{A}_\parallel=\frac{2(m_{B_q}+m_V)A_1(q^2)}{\sqrt 2}\,,\qquad\qquad
\mathcal{A}_\perp=-\frac{4m_{B_q}V(q^2)|\vec p_V|}{\sqrt{2}(m_{B_q}+m_V)}\,,\nonumber\\
&&\mathcal{A}_t=\frac{2m_{B_q}|\vec p_V|A_0(q^2)}{\sqrt {q^2}}\,,\qquad\qquad
\mathcal{A}_P=-\frac{2m_{B_q}|\vec p_V|A_0(q^2)}{(m_b(\mu)+m_c(\mu))}\,.
\end{eqnarray}
 
For the details of the helicity amplitudes, $B$ meson decay constant, and the $B_q \to (P, V)$ meson 
transition form factors, we refer to Refs.~\cite{dutta, Bhol:2014jta}.

To study the possibility of correlation in $\tau$ decays, we follow Ref.~\cite{Nandi:2016wlp} and 
define new observables $R_D^{\tau}$ and $R^{\tau}_{D^{\ast}}$ as
\begin{eqnarray}
&&R^{\tau}_{D} = \frac{R_D}{\mathcal B(B \to \tau\nu)}\,, \qquad\qquad
R^{\tau}_{D^{\ast}} = \frac{R_{D^{\ast}}}{\mathcal B(B \to \tau\nu)}\,.
\end{eqnarray}
The $\tau$ detection and identification systematics that are present in both $B \to D\,\tau\,\nu$ and $B \to D^{\ast}\,\tau\,\nu$ decays 
may get cancelled in these new ratios.
Semileptonic $B_s$ decays to $D_s\,\tau\,\nu$ and $D^{\ast}_s\,\tau\,\nu$ and $B$ decays to $\pi\,\tau\,\nu$ are also mediated via
$b \to (u,\,c)\,\tau\,\nu$ quark level transition processes and, in principle, are subject to NP. In this context, we also define ratio 
of branching ratios in these decay modes similar to $B \to (D,\,D^{\ast})\,\tau\,\nu$
decays. Those are 
\begin{eqnarray}
&&R_{\pi} = \frac{\mathcal B(B \to \pi\tau\nu)}{\mathcal B(B \to \pi\,l\,\nu)}\,,\qquad\qquad
R_{D_s} = \frac{\mathcal B(\bar B_s \to D_s\tau^- \bar{\nu}_\tau)}{\mathcal B(\bar B_s \to D_s\, l^- \bar{\nu}_l)}\,, \qquad\qquad
R_{D^{\ast}_s}= \frac{\mathcal B(\bar B_s \to {D^{\ast}_s}\tau^- \bar{\nu}_\tau)}{\mathcal B(\bar B_s \to {D^{\ast}_s}\,l^- \bar{\nu}_l)}\,, 
\nonumber \\
&&R^{\tau}_{\pi} = \frac{\mathcal B(B \to \pi\tau\nu)}{\mathcal B(B \to \tau\nu)}\,, \qquad\qquad
R^{\tau}_{D_s} = \frac{R_{D_s}}{\mathcal B(B \to \tau\nu)}\,, \qquad\qquad
R^{\tau}_{D^{\ast}_s} = \frac{R_{D^{\ast}_s}}{\mathcal B(B \to \tau\nu)}\,.
\end{eqnarray}
We want to mention that although $R_{\pi}$, $R_{D_s}$, $R_{D^{\ast}_s}$, and $R^{\tau}_{\pi}$ do not depend on CKM matrix
elements $V_{ub}$ and $V_{cb}$, but the newly constructed ratios $R^{\tau}_{D}$, $R^{\tau}_{D^{\ast}}$, $R^{\tau}_{D_s}$ and 
$R^{\tau}_{D^{\ast}_s}$ do depend on the CKM matrix element $V_{ub}$.

We wish to see the effect of various NP couplings on these observables in a model independent way.
There are two types of uncertainties in theoretical calculation of the observables. First kind of  uncertainties may come from the
very well known input parameters such as quark
masses, meson masses, and the mean life time of mesons. We ignore such uncertainties
as they are not important for our
analysis. Second kind of uncertainties may arise due to not very well known parameters such as CKM matrix elements, meson decay constants,
and the meson to meson transition form factors. In order to gauge the effect of above mentioned uncertainties on various
observables, we use a random number generator and perform a random scan of all the theoretical inputs such as CKM matrix elements, 
meson decay constants, and the meson to meson transition form factors. We vary all the theoretical inputs within $2\sigma$ from their
central values in our random scan. The allowed NP parameter space is obtained by imposing $2\sigma$ constraints coming from BABAR and
BELLE measured values of the ratio of branching ratios $R_D$, $R_{D^{\ast}}$, and $\mathcal B(B \to \tau\nu)$. We also use $2\sigma$
constraints coming from the estimated values of the newly constructed ratios $R_D^{\tau}$ and $R_{D^{\ast}}^{\tau}$ to 
explore various NP couplings.
We now proceed to discuss the results of our analysis.
\section{Numerical calculations}
\label{rd}
For definiteness, let us first give the details of the input parameters that are used for the theoretical computation of all the 
observables. For the quark mass, meson mass, and the meson life time, we use the following input parameters from Ref.~\cite{Olive:2016xmw}.
\begin{eqnarray}
\label{inputs}
&&m_b(m_b) = 4.18\,{\rm GeV}\,,\qquad\qquad
m_c(m_b)=0.91\,{\rm GeV}\,,\qquad\qquad
m_{\pi} = 0.13957\,{\rm GeV}\,\nonumber \\
&&m_{B^{-}} = 5.27925\,{\rm GeV}\,,\qquad\qquad
m_{B^{0}} = 5.27955\,{\rm GeV}\,,\qquad\qquad
m_{B_{s}} = 5.36677\,{\rm GeV}\,,\nonumber \\
&&m_{D^0} = 1.86486\,{\rm GeV}\,,\qquad\qquad
m_{D^{\ast\,0}} = 2.00698\,{\rm GeV}\,,\qquad\qquad
m_{D_s^+} = 1.9685\,{\rm GeV}\,,\nonumber\\
&&  m_{{D_s^*}^+} = 2.1123\,{\rm GeV}\,,\qquad\qquad
\tau_{B^0} = 1.519\times 10^{-12}\,{\rm Sec}\,,\qquad\qquad
\tau_{B^-} = 1.641\times 10^{-12}\,{\rm Sec}\,,\nonumber\\
&& \tau_{B_s} = 1.516\times 10^{-12}\,{\rm Sec}\,
\end{eqnarray}
Similarly, for the CKM matrix elements, meson decay constant, and meson to meson transition form factors, we use the inputs that 
are tabulated in Table~\ref{tab1}. We refer to Refs.~\cite{dutta, Bhol:2014jta} for a detailed discussion on various form 
factor calculation.
\begin{table}[htdp]
\begin{center}
\begin{tabular}{|c|c|c|c|c|}
\hline
\multicolumn {2}{|c|}{CKM matrix Elements:} & \multicolumn {3}{|c|}{Meson Decay constants~(in GeV)
:} \\[0.2cm]
\hline
$ |V_{ub}| $ (Exclusive) & $(3.61 \pm 0.32 ) \times 10^{-3} $~\cite{pdg} & $ f_B $ & 
\multicolumn{2}{|c|} {$  0.1906 \pm 0.0047 $~\cite{Bazavov:2011aa, Na:2012kp, latticeavg}} \\[0.2cm]
$ |V_{cb}| $ (Average) & $ (40.9 \pm 1.1 ) \times 10^{-3} $~\cite{pdg} &$  $ & \multicolumn{2}{|c|} {$ $} \\[0.2cm]
\hline
\multicolumn{2}{|c|}{Inputs for $(B \to \pi)$ Form Factors:} & \multicolumn{3}{|c|}{Inputs for $(B \to D^{\ast})$ Form Factors:} \\[0.2cm]
\hline
$ F_{+}(0)=F_{0}(0)$ & $ 0.281\pm0.028$ ~\cite{Khodjamirian} & $ h_{A_1}(1)|V_{cb}| $ & \multicolumn{2}{|c|} {$ (34.6\pm1.02)\times 
10^{-3} $ ~\cite{Dungel:2010uk}} \\[0.2cm]
$ b_1 $ & $ -1.62\pm 0.70 $ ~\cite{Khodjamirian} & $ \rho_1^2 $ & \multicolumn{2}{|c|} {$ 1.214\pm 0.035$ ~\cite{Dungel:2010uk}} \\[0.2cm]
$ b_1^0 $ & $ -3.98\pm 0.97 $ ~\cite{Khodjamirian} & $ R_1(1) $ & \multicolumn{2}{|c|} {$ 1.401\pm 0.038$ ~\cite{Dungel:2010uk}} \\[0.2cm]
\cline{1-2}
\cline{1-2}
\multicolumn{2}{|c|}{Inputs for $(B \to D)$ Form Factors:} & $ R_2(1) $ & \multicolumn{2}{|c|} {$ 0.864\pm 0.025$ ~\cite{Dungel:2010uk}} 
\\[0.2cm]
\cline{1-2}
$ V_1(1)|V_{cb}| $ & $ (43.0\pm2.36)\times 10^{-3} $ ~\cite{Aubert:2009ac} & $ R_0(1) $ & \multicolumn{2}{|c|} 
{$ 1.14\pm 0.114$ ~\cite{Fajfer2}} \\[0.2cm]
$ \rho_1^2 $ & $ 1.20\pm 0.098$ ~\cite{Aubert:2009ac} & $ $  & \multicolumn{2}{|c|} {$ $} \\[0.2cm]
\hline
\multicolumn{5}{|c|}{Inputs for $(B_s \to D_s)$ Form Factors:~\cite{Faustov}} \\[0.2cm]
\hline
  $ $ &  \multicolumn{2}{|c|} {$ F_+ $} & \multicolumn{2}{|c|} {$ F_0 $}  \\[0.2cm]
\hline
$ F(0) $ & \multicolumn{2}{|c|} { $0.74\pm0.02 $} & \multicolumn{2}{|c|} {$0.74\pm0.02 $}  \\[0.2cm]
$ \sigma_1 $ & \multicolumn{2}{|c|} {$ 0.20\pm0.02 $} & \multicolumn{2}{|c|}{$ 0.430\pm0.043 $} \\[0.2cm]
$ \sigma_2 $ & \multicolumn{2}{|c|} {$ -0.461\pm0.0461 $} & \multicolumn{2}{|c|}{$ -0.464\pm0.0464 $} \\[0.2cm]
\hline
\multicolumn{5}{|c|}{Inputs for $(B_s \to D_s^*)$ Form Factors:~\cite{Faustov}} \\[0.2cm]
\hline
  $ $ &  $V $ & $A_0$ & $ A_1$ & $A_2$ \\[0.2cm]
\hline
$ F(0) $ & {$ 0.95\pm0.02 $} & {$ 0.67\pm0.01 $} & $ 0.70\pm0.01 $ & $ 0.75\pm0.02 $ \\[0.2cm]
$ \sigma_1 $ & {$ 0.372\pm0.0372 $} & {$ 0.350\pm0.035 $} & $ 0.463\pm0.0463 $ & $ 1.04\pm0.104 $ \\[0.2cm]
$ \sigma_2 $ & {$ -0.561\pm.0561 $} & {$ -0.60\pm0.06 $} & $ -0.510\pm0.051 $ & $ -0.07\pm0.007 $ \\[0.2cm]
\hline
\end{tabular}
\end{center}
\caption{\footnotesize{Theory Input parameters}}
\label{tab1}
\end{table}
The uncertainties associated with both the theory and experimental input parameters are 
added in quadrature and tabulated in Table~\ref{tab1} and Table~\ref{tab2}. The SM prediction for all the observables 
are reported in Table.~\ref{tab3}. Central values of all the observables are obtained by using the central values of all the 
input parameters from Eq.~(\ref{inputs}) and 
from Table.~\ref{tab1}.
The $1\sigma$ range in each observable, reported in Table.~\ref{tab3}, is obtained by performing a random scan of all the theory 
inputs such as
$B_q$ meson decay constants, $B_q \to (P,V)$ transition form factors and the CKM matrix 
elements $|V_{qb}|$ within $1\sigma$ of their central values.

\begin{table}[htdp]
\parbox{.45\linewidth}{
\centering
\begin{tabular}{|c|c|c|}
\hline
\multicolumn{3}{|c|}{Ratio of branching ratios:} \\[0.2cm]
\hline
$ {\rm BABAR}$ & $ \mathcal B(B \to \tau\nu) $ & $(1.83 \pm 0.52)\times 10^{-4} $ ~\cite{Lees:2012ju} \\[0.2cm]
$ {\rm BELLE} $ & $ \mathcal B(B \to \tau\nu) $ & $(1.25 \pm 0.28)\times 10^{-4} $ ~\cite{Kronenbitter:2015kls} \\[0.2cm]
$ {\rm BABAR}$ & $ R_D $ & $0.440 \pm 0.072 $ ~\cite{Lees:2012xj, Lees:2013uzd} \\[0.2cm]
$ {\rm BABAR}$ & $ R_{D^{\ast}} $ & $0.332 \pm 0.030 $ ~\cite{Lees:2012xj, Lees:2013uzd} \\[0.2cm]
$ {\rm BELLE} $ & $ R_D $ & $0.375 \pm 0.069 $ ~\cite{Huschle:2015rga} \\[0.2cm]
$ {\rm BELLE}$ & $  R_{D^{\ast}} $ & $0.293 \pm 0.0.041 $ ~\cite{Huschle:2015rga} \\[0.2cm]
$ {\rm BABAR}$ & $ R_D^\tau $ & $ (2.404 \pm 0.838)\times 10^3 $ ~\cite{Nandi:2016wlp} \\[0.2cm]
$ {\rm BABAR}$ & $ R^\tau_{D^{\ast}} $ & $(1.814 \pm 0.582)\times 10^3 $ ~\cite{Nandi:2016wlp} \\[0.2cm]
$ {\rm BELLE} $ & $ R_D^\tau $ & $ (3.0 \pm 1.1)\times 10^3 $ ~\cite{Nandi:2016wlp} \\[0.2cm]
$ {\rm BELLE} $ & $ R^\tau_{D^{\ast}} $ & $ (2.344 \pm 0.799)\times 10^3 $ ~\cite{Nandi:2016wlp} \\[0.2cm]
\hline
\end{tabular}
\caption{\footnotesize{Experimental Input parameters}}
\label{tab2}
}
\hfill
\parbox{.45\linewidth}{
\centering
\begin{tabular}{|c|c|c|}
\hline
$ $ & Central value & $1\sigma$ range\\
\hline
$R^\tau_{D} $ & $3.737\times 10^{3}$ & $(2.889,\,4.919)\times 10^{3}$ \\[0.2cm]
$R^\tau_{D^{\ast}}$ & $3.022\times 10^{3}$ & $(2.375,\,3.918)\times 10^{3}$ \\[0.2cm]
$R^\tau_\pi$ & $1.33$ & $(0.847, 2.015)$  \\[0.2cm]
$R_\pi$ & $0.698$ & $(0.654,0.764)$  \\[0.2cm]
$R^\tau_{D_s}$ & $3.270\times 10^{3}$ & $(2.499, \, 4.396)\times 10^{3}$ \\[0.2cm]
$R^\tau_{D_s^{\ast}}$ & $2.881\times 10^{3}$ & $(2.295,\,3.687)\times 10^{3}$ \\[0.2cm]
$R_{D_s}$ & $0.274$ & $(0.255,0.295)$ \\[0.2cm]
$R_{D_s^{\ast}}$ & $0.241$ & $(0.236,0.246)$ \\[0.2cm]
\hline
\end{tabular}
\caption{\footnotesize{SM prediction of various observables.}}
\label{tab3}
}
\end{table}
Our main aim is to study NP effects on various new observables such as $R_{\pi}$, $R_{D_s}$, $R_{D^{\ast}_s}$,
$R^\tau_{D}$, $R^\tau_{D^{\ast}}$, $R^\tau_{\pi}$, $R^\tau_{D_s}$, $R^\tau_{D_s^{\ast}}$ in a model 
independent way. We consider four different NP scenarios. First, we use $2\sigma$ experimental 
constraint coming from the BABAR and BELLE measured values of the ratio of 
branching ratios $R_D$ and $R_{D^{\ast}}$, and $\mathcal B(B \to \tau\nu)$. Second,
we put additional constraint coming from the estimated values of $R_D^{\tau}$ and $R_{D^{\ast}}^{\tau}$. 
The observables $R^\tau_{D}$ and $R^\tau_{D^{\ast}}$ are ratios obtained by dividing the ratio of branching ratios 
$R_D$ and $R_{D^{\ast}}$ by branching ratio $\mathcal B(B \to \tau\nu)$. Hence the NP effect will be cancelled 
to a large extent in these ratios. Moreover, the estimated values of these new ratios are consistent with the SM prediction. 
Although, it does not necessarily rule out the presence of NP, it may, however, constrain the NP parameter 
space even further. Again, the $\tau$ detection systematics will also largely cancel in these ratios.
Because of the presence of $V_{ub}$ in these ratios, the estimated errors on both these observables
are rather large. However, this could be reduced once more precise data on $V_{ub}$ is available.
In view of the anticipated improved precision in the measurement of $V_{ub}$, we impose $2\sigma$ experimental constraint 
coming from the estimated values of $R^\tau_{D}$ and $R^\tau_{D^{\ast}}$ in addition to the constraints coming from 
$R_D$, $R_{D^{\ast}}$, and $\mathcal B(B \to \tau\nu)$ to explore various NP couplings.
All the NP parameters are 
considered to be real for our analysis. We also assume that only the third generation 
leptons get contributions from the NP couplings in the $b\to(u,\,c)\,l\nu $ processes and 
for $l=e^-,\mu^-$ cases, NP is absent. We next discuss the effect of various NP couplings after imposing constraints from BABAR
and BELLE measurements.
 
\subsection{BABAR constraint}
\label{babar}
\begin{figure}
\begin{center}
\includegraphics[width=6cm,height=4cm]{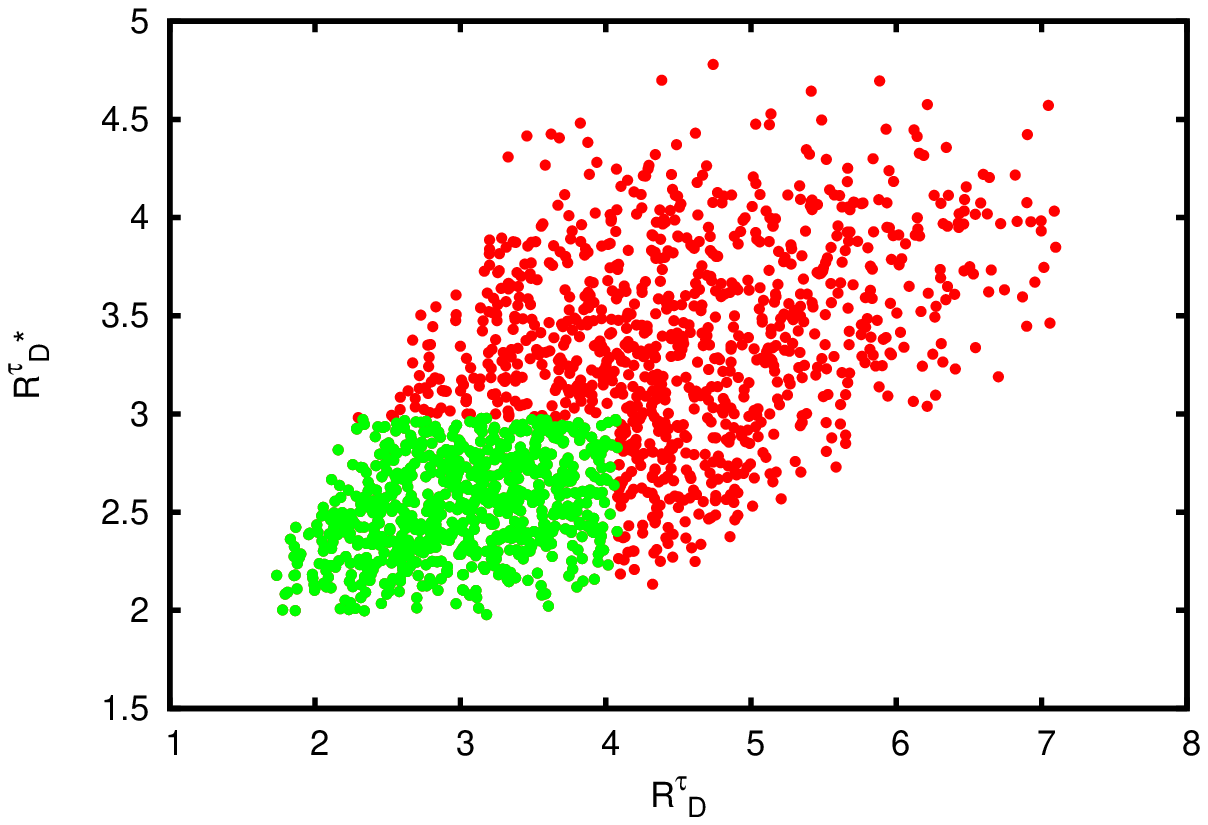}
\includegraphics[width=6cm,height=4cm]{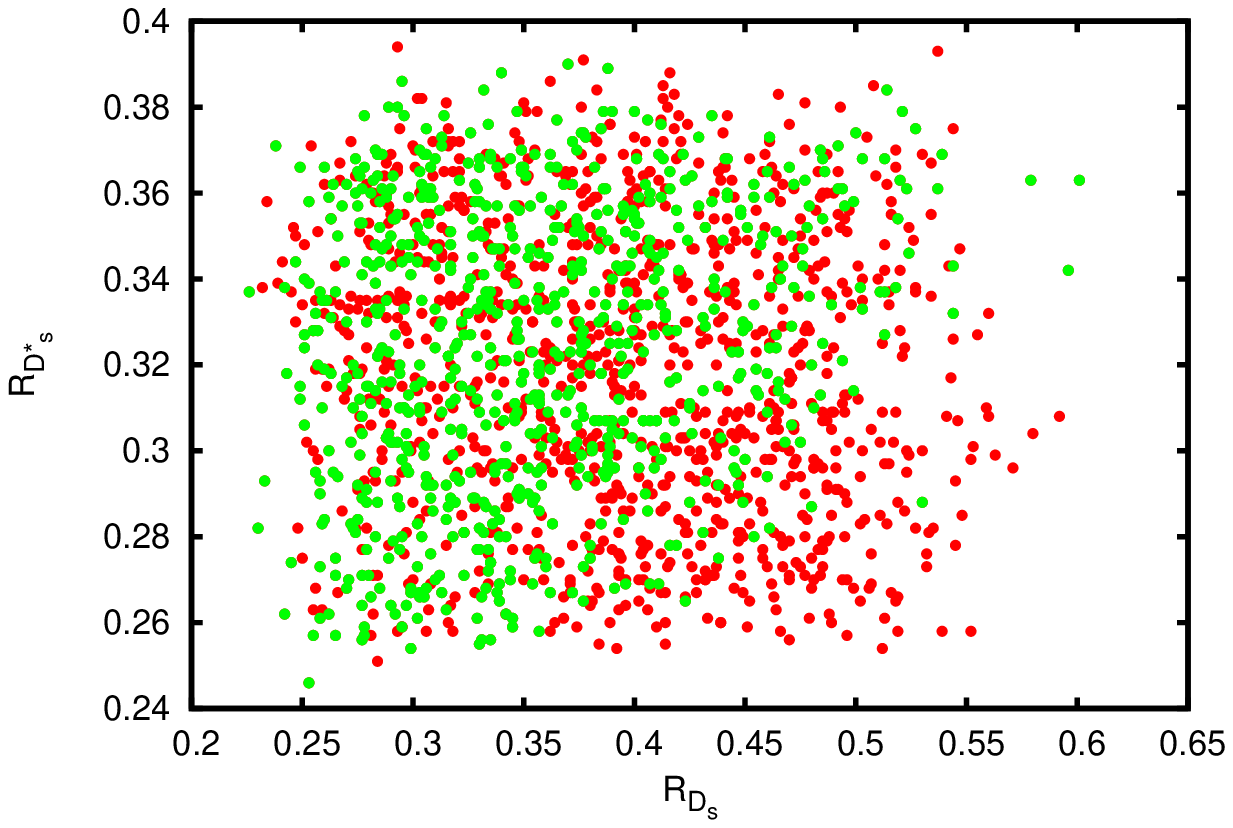}
\includegraphics[width=6cm,height=4cm]{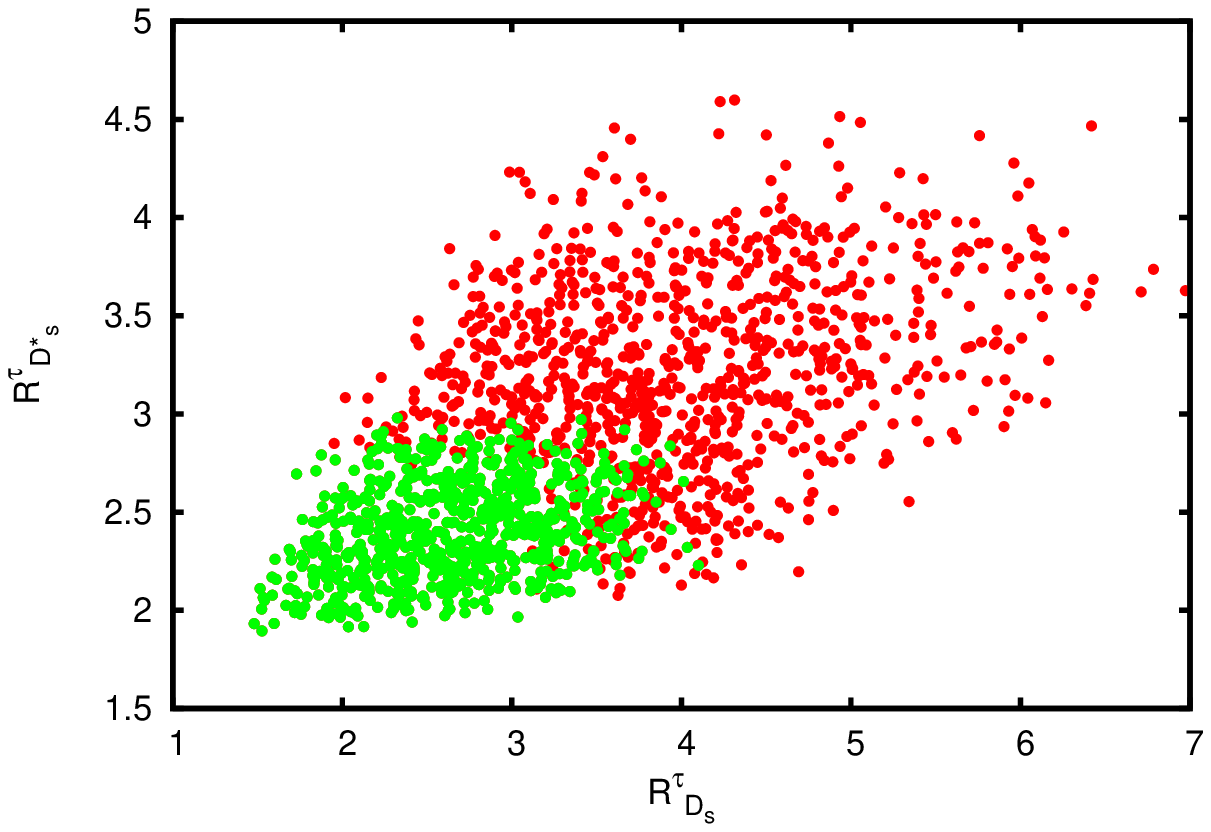}
\includegraphics[width=6cm,height=4cm]{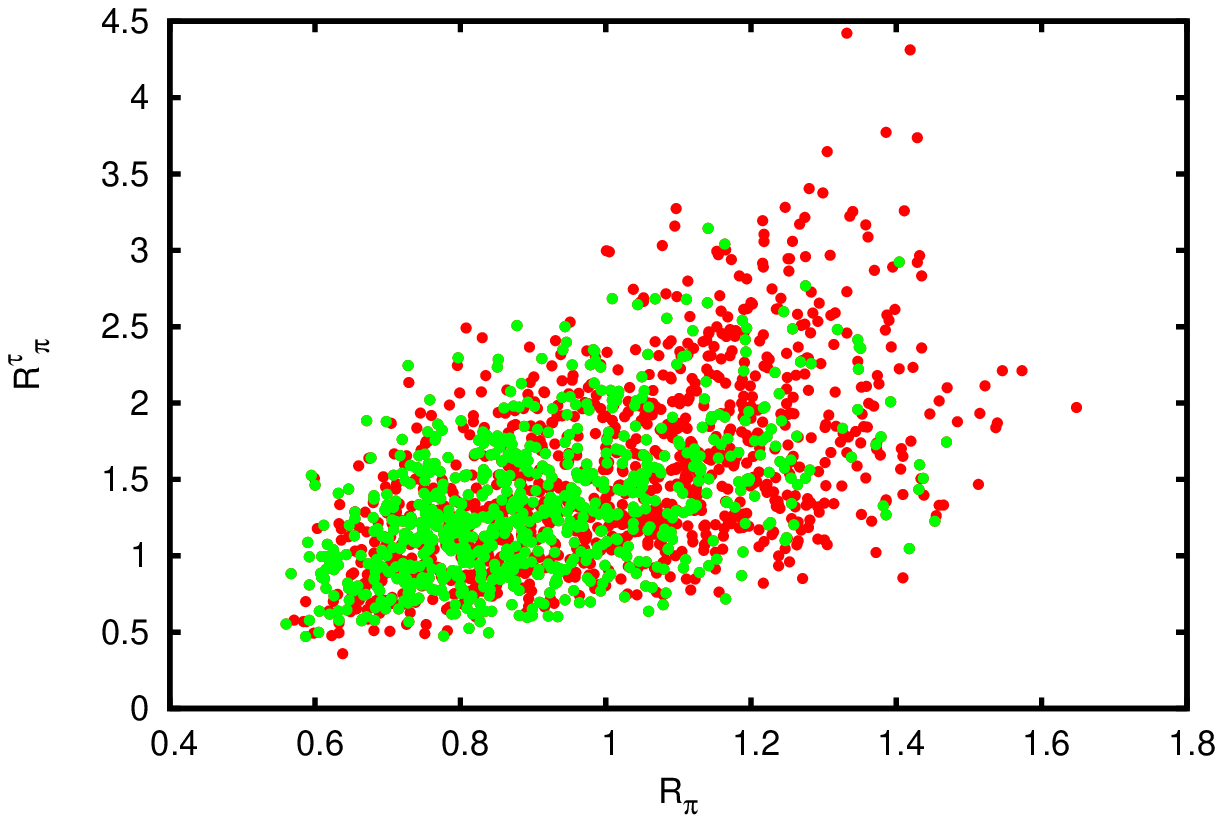}
\end{center}
\caption{\footnotesize{Allowed ranges in various observables with $V_L$ and $V_R$ type NP couplings once the BABAR $2\sigma$ experimental 
constraint is imposed. The dark~(red) region corresponds to the allowed ranges once $R_D$, $R_{D^{\ast}}$, and $\mathcal
B(B \to \tau\nu)$ constraint is imposed, whereas, the light~(green) region corresponds to the allowed ranges of the observables
once additional $2\sigma$ constraint from $R_D^{\tau}$ and $R_{D^{\ast}}^{\tau}$ is imposed.}}
\label{obs_bab_np1}
\end{figure}

We consider four different NP scenarios for our analysis.
In the first scenario, we vary new vector couplings $V_L$ and $V_R$ and consider all other NP couplings to 
be zero. First, we impose $2\sigma$ experimental constraint coming from BABAR measured values of the 
ratio of branching ratios $R_D$, $R_{D^{\ast}}$, and $\mathcal B(B \to \tau\nu)$  
to constrain the new vector type couplings $(V_L,\,V_R)$. Second, we impose additional $2\sigma$ constraint coming from the 
estimated values of 
$R_D^{\tau}$ and $R_{D^{\ast}}^{\tau}$ to see whether it is possible to constrain the NP parameter space even further. 
In Fig.~\ref{obs_bab_np1} we show the NP effect on various 
observables after imposing the $2\sigma$ experimental constraint coming from the
BABAR measured values. 
The allowed ranges in each observable are tabulated in Table.~\ref{tab4}.
\begin{table}[htdp]
\vspace*{-0.1cm}
\begin{center}
\begin{tabular}{||c|c|c||c|c|c||}
\hline
Observable &Column I &Column II &Observable & Column I& Column II\\
\hline
\hline
$R_D^{\tau}(\times 10^3)$ &$(1.736, 7.097)$ &$(1.736, 4.080)$ &$R_{\pi}^{\tau}$ &$(0.359, 4.422)$ &$(0.472, 3.144)$ \\[0.2cm]
\hline
$R_{D^{\ast}}^{\tau}(\times 10^3)$ &$(1.978, 4.780)$ &$(1.978, 2.978)$ &$R_{\pi}$ &$(0.560,1.648)$ & $(0.560, 1.469)$\\[0.2cm]
\hline
$R_{D_s}^{\tau}(\times 10^3)$ &$(1.480, 6.973)$ &$(1.480, 4.101)$ &$R_{D_s}$ &$(0.226, 0.601)$ &$(0.226, 0.601)$ \\[0.2cm]
\hline
$R_{{D^{\ast}_s}}^{\tau}(\times 10^3)$ &$(1.895, 4.599)$ &$(1.895, 2.980)$ &$R_{D^{\ast}_s}$ &$(0.246, 0.394)$ &$(0.246, 0.390)$ \\[0.2cm]
\hline
\hline
\end{tabular}
\end{center}
\caption{\footnotesize{Allowed ranges in various observables with $(V_L,\,V_R)$ NP couplings. The ranges reported 
in Column I represent the allowed values of each observable once constraints coming from BABAR
measured values of $R_D$, $R_{D^{\ast}}$, and $\mathcal B(B \to \tau\nu)$ are imposed, whereas, the ranges in Column II represent 
the allowed values once additional $2\sigma$ constraints from $R_D^{\tau}$ and $R_{D^{\ast}}^{\tau}$ are imposed. }}
\label{tab4}
\end{table}
We find significant deviation of all the observables from SM expectation in this scenario.
It is clear that we can constrain the NP parameter space 
even further by imposing constraints coming from $R_D^{\tau}$ and $R_{D^{\ast}}^{\tau}$. To illustrate this point, we show 
with green dots the NP effect on various observables once additional $2\sigma$ constraints coming from
$R_D^{\tau}$ and $R_{D^{\ast}}^{\tau}$ are imposed. It is observed that the allowed ranges in $R_D^{\tau}$, $R_{D^{\ast}}^{\tau}$, 
$R_{D_s}^{\tau}$, $R_{D^{\ast}_s}^{\tau}$, and $R_{\pi}^{\tau}$ are considerably reduced whereas there are no or very little changes 
in $R_{\pi}$, $R_{D_s}$, 
and $R_{D^{\ast}_s}$ allowed ranges once the additional $2\sigma$ constraint from $R_D^{\tau}$ and $R_{D^{\ast}}^{\tau}$ are imposed.
We want to emphasize that since the new observables, $R^\tau_{D_{(s)}^{(\ast)}}$ and $R_{\pi}^{\tau}$ are ratios obtained by
normalizing $R_{D_{(s)}^{(*)}}$ and $\mathcal B(B \to \pi\tau\nu)$ with the branching ratio $\mathcal B(B \to \tau\nu)$,
there must be some cancellation of NP effects. However, the NP effect can not
be completely eliminated. NP effect will not be present in these new ratios if only $V_L$ type NP
couplings are present. In that case $G_V = G_A$ and the contribution coming from the NP couplings will cancel
in $R^\tau_{D_{(s)}^{(\ast)}}$ and $R_{\pi}^{\tau}$.

\begin{figure}
\begin{center}
\includegraphics[width=6cm,height=4cm]{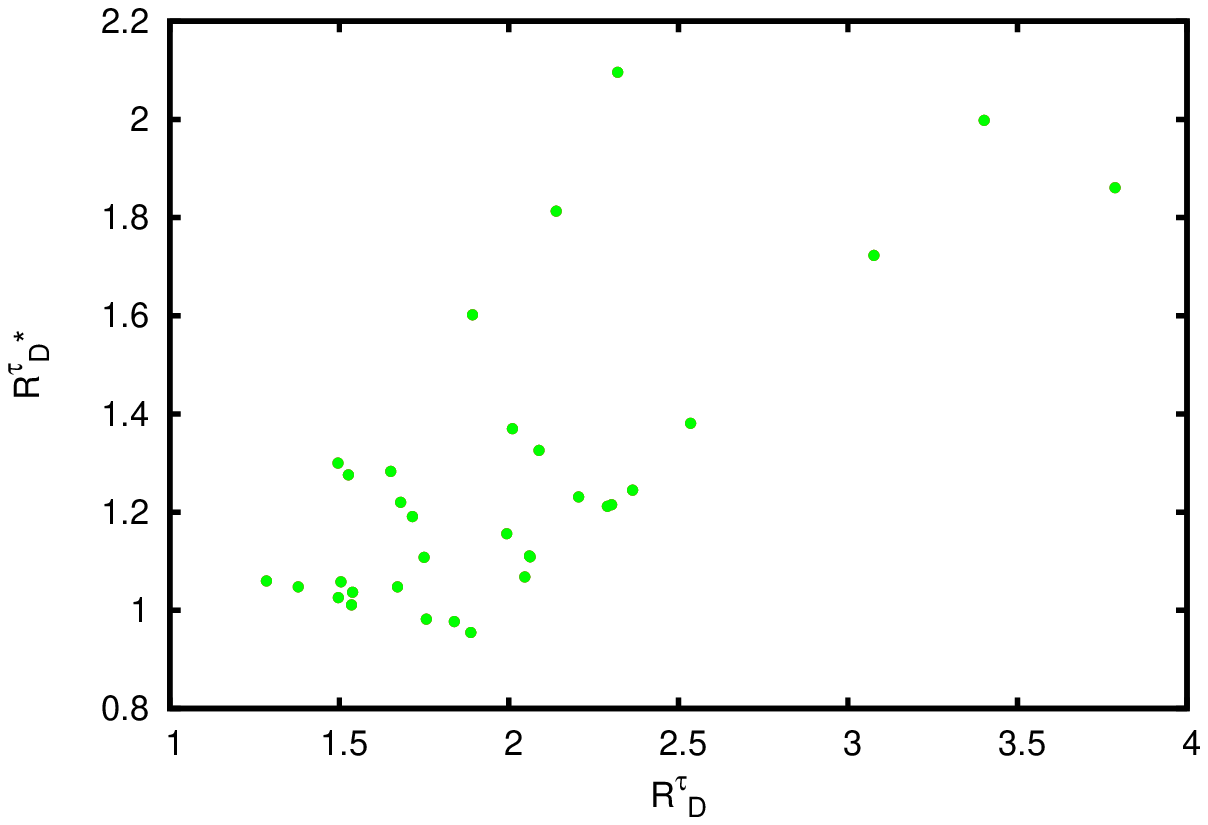}
\includegraphics[width=6cm,height=4cm]{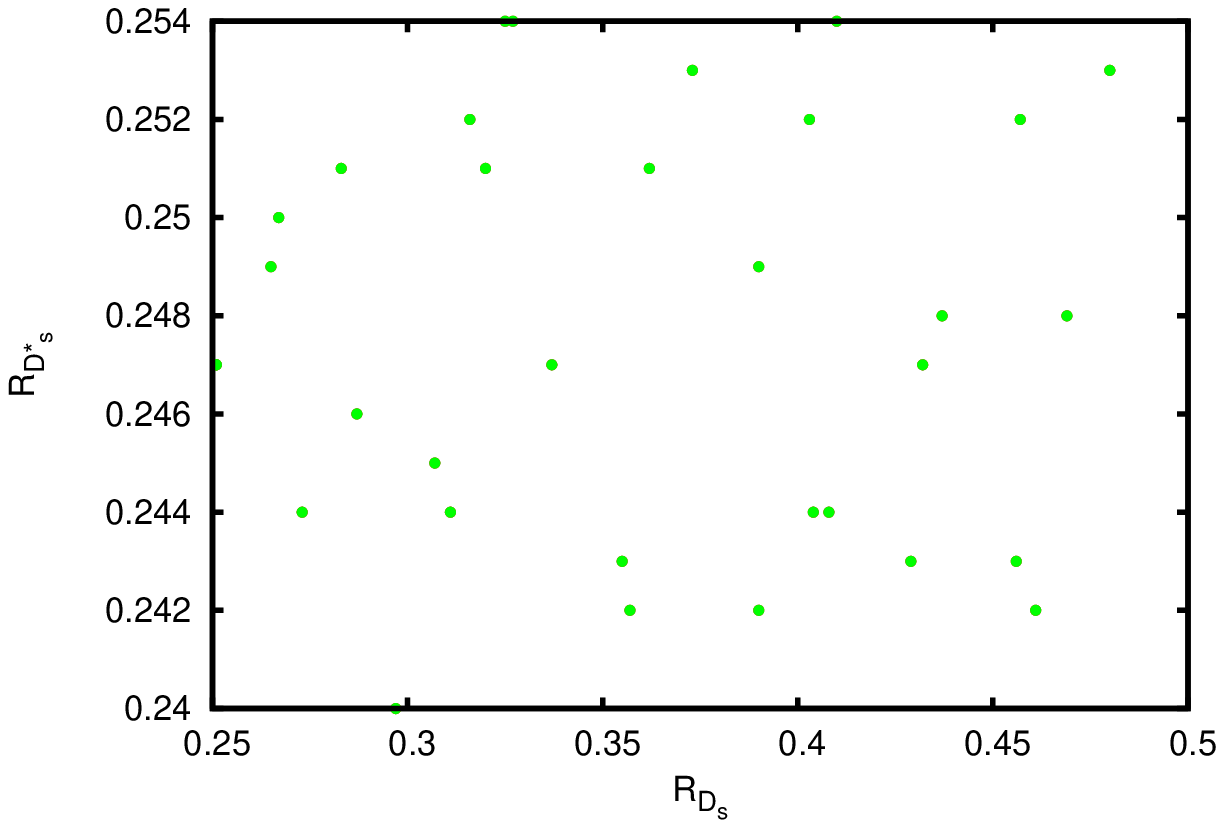}
\includegraphics[width=6cm,height=4cm]{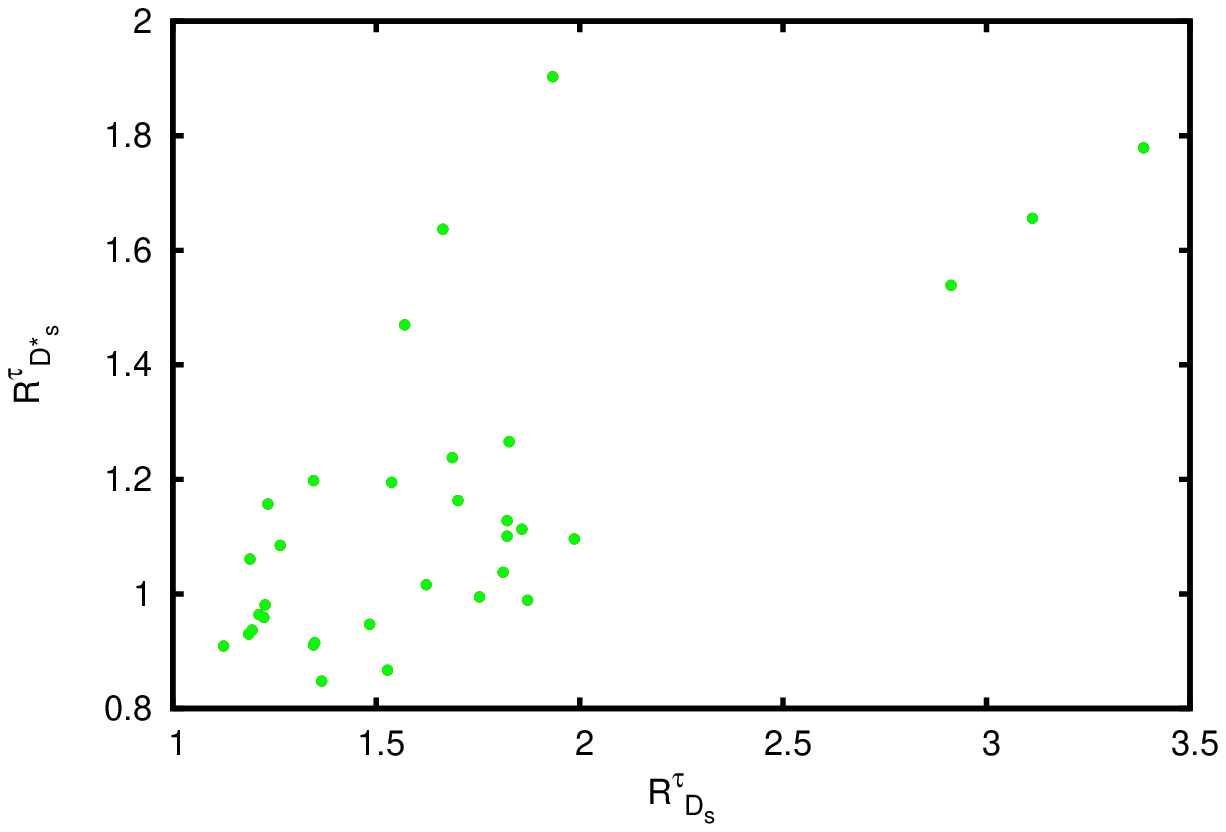}
\includegraphics[width=6cm,height=4cm]{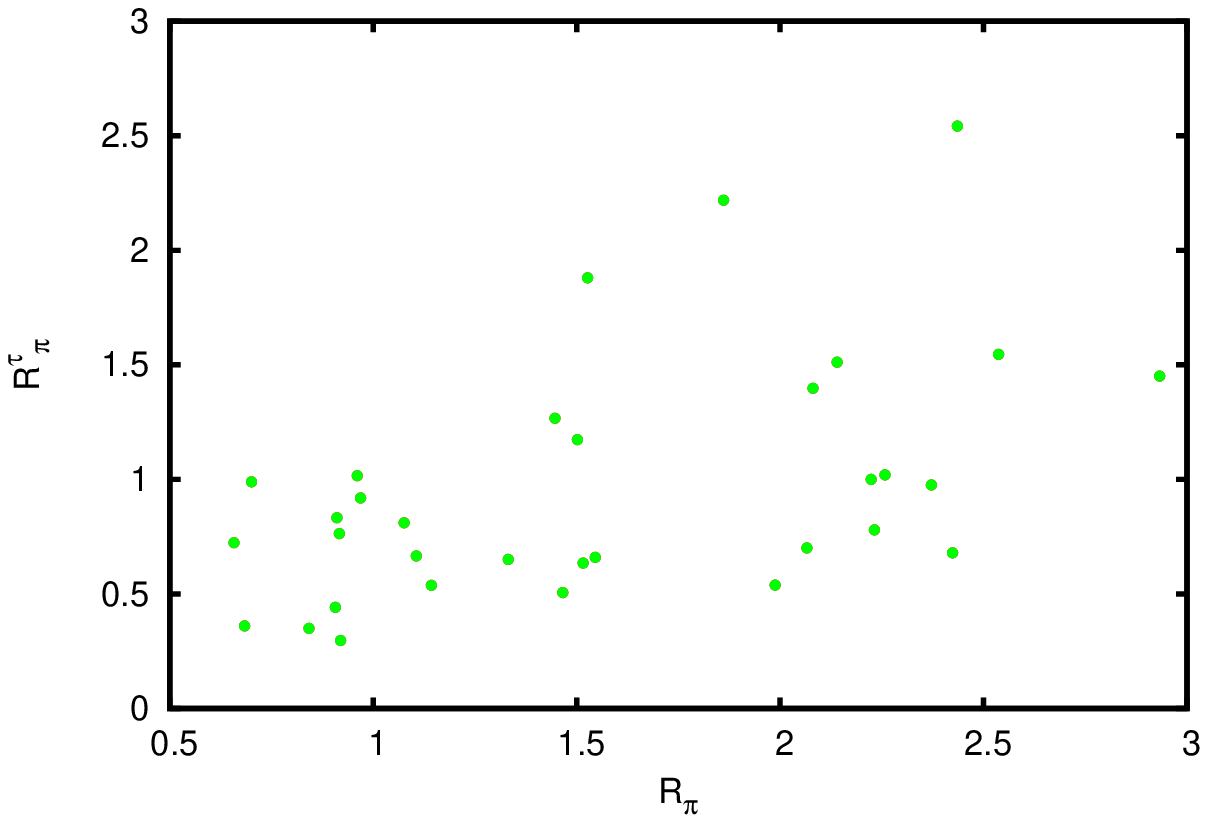}
\end{center}
\caption{\footnotesize{Allowed ranges in various observables with $S_L$ and $S_R$ type NP couplings once the BABAR $2\sigma$ experimental 
constraint is imposed. Allowed range obtained by imposing $2\sigma$ constraint coming from $R_D$, $R_{D^{\ast}}$,
and $\mathcal B(B \to \tau\nu)$ overlaps with the allowed range once additional $2\sigma$
constraint from $R_D^{\tau}$ and $R_{D^{\ast}}^{\tau}$ is imposed.}} 
\label{obs_bab_np2}
\end{figure}

In the second scenario, we study the impact of new scalar couplings $S_L$ and $S_R$ on various observables keeping all other NP 
couplings to be zero. 
The effect of $S_L$ and $S_R$ type NP couplings on various observables are shown in Fig.~\ref{obs_bab_np2} once the
$2\sigma$ experimental constraints coming from BABAR measured values are imposed. Significant deviation from the SM expectation 
is observed in this scenario. Again, putting additional $2\sigma$ constraints from  $R_D^{\tau}$ and $R_{D^{\ast}}^{\tau}$ do not 
seem to affect any of the observables.
The allowed ranges in each observable are tabulated in Table.~\ref{tab5}.
\begin{table}[htdp]
\vspace*{-0.1cm}
\begin{center}
\begin{tabular}{||c|c|c||c|c|c||}
\hline
Observable & Column I  & Column II &Observable & Column I& Column II\\
\hline
\hline
$R_D^{\tau}(\times 10^3)$ &$(1.285, 3.788)$ &$(1.285, 3.788)$ &$R_{\pi}^{\tau}$ &$(0.297, 2.542)$ &$(0.297, 2.542)$ \\[0.2cm]
\hline
$R_{D^{\ast}}^{\tau}(\times 10^3)$ &$(0.955, 2.096)$ &$(0.955, 2.096)$ &$R_{\pi}$ &$(0.658, 2.933)$ &$(0.658, 2.933)$ \\[0.2cm]
\hline
$R_{D_s}^{\tau}(\times 10^3)$ &$(1.125, 3.386)$ &$(1.125, 3.386)$ &$R_{D_s}$ &$(0.251, 0.480)$ &$(0.251, 0.480)$ \\[0.2cm]
\hline
$R_{{D^{\ast}_s}}^{\tau}(\times 10^3)$ &$(0.848, 1.903)$ &$(0.848, 1.903)$ &$R_{D^{\ast}_s}$ &$(0.240, 0.254)$ &$(0.240, 0.254)$ \\[0.2cm]
\hline
\hline
\end{tabular}
\end{center}
\caption{\footnotesize{Allowed ranges in various observables with $(S_L,\, S_R)$ NP couplings. The ranges reported in Column I represent 
the allowed values of each observable once constraints coming from BABAR
measured values of $R_D$, $R_{D^{\ast}}$, and $\mathcal B(B \to \tau\nu)$ are imposed. Column II represents the allowed range in
each observable once additional $2\sigma$ constraints from $R_D^{\tau}$
and $R_{D^{\ast}}^{\tau}$ are imposed.}}
\label{tab5}
\end{table}

\begin{figure}
\begin{center}
\includegraphics[width=6cm,height=4cm]{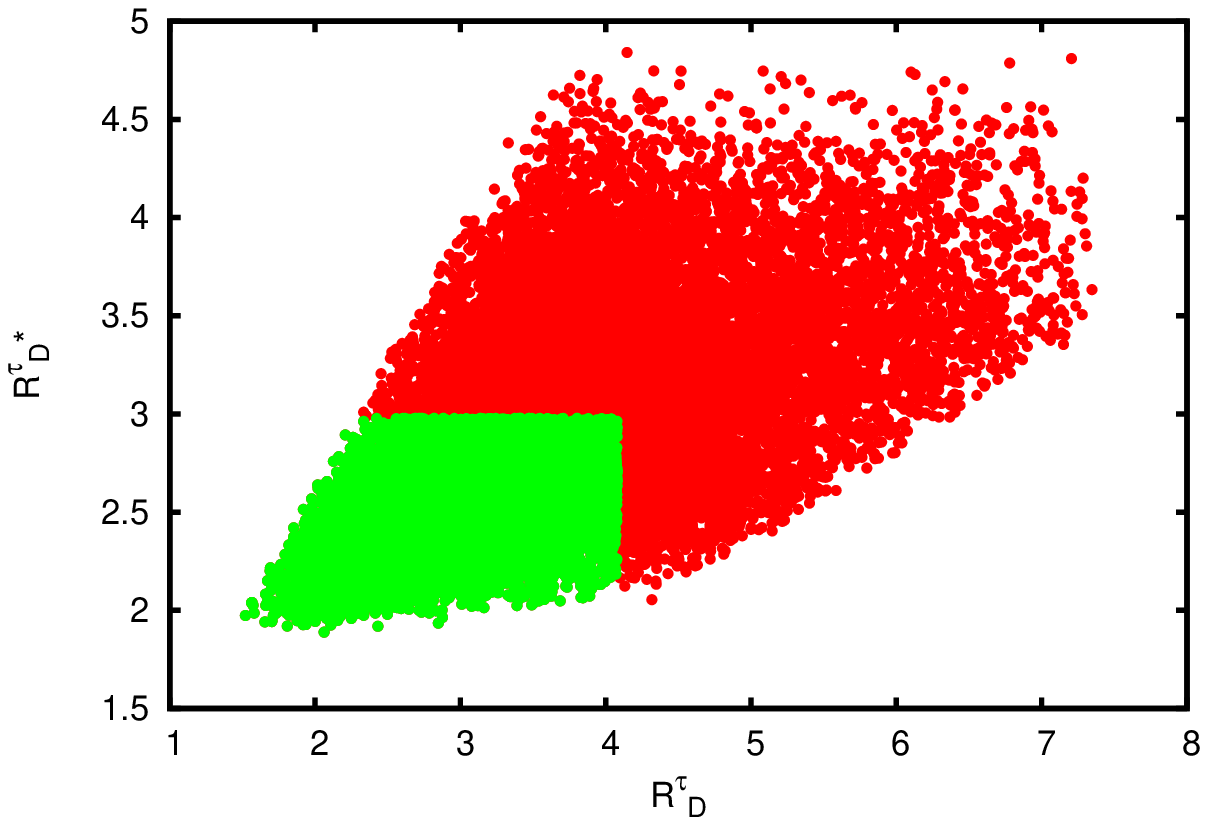}
\includegraphics[width=6cm,height=4cm]{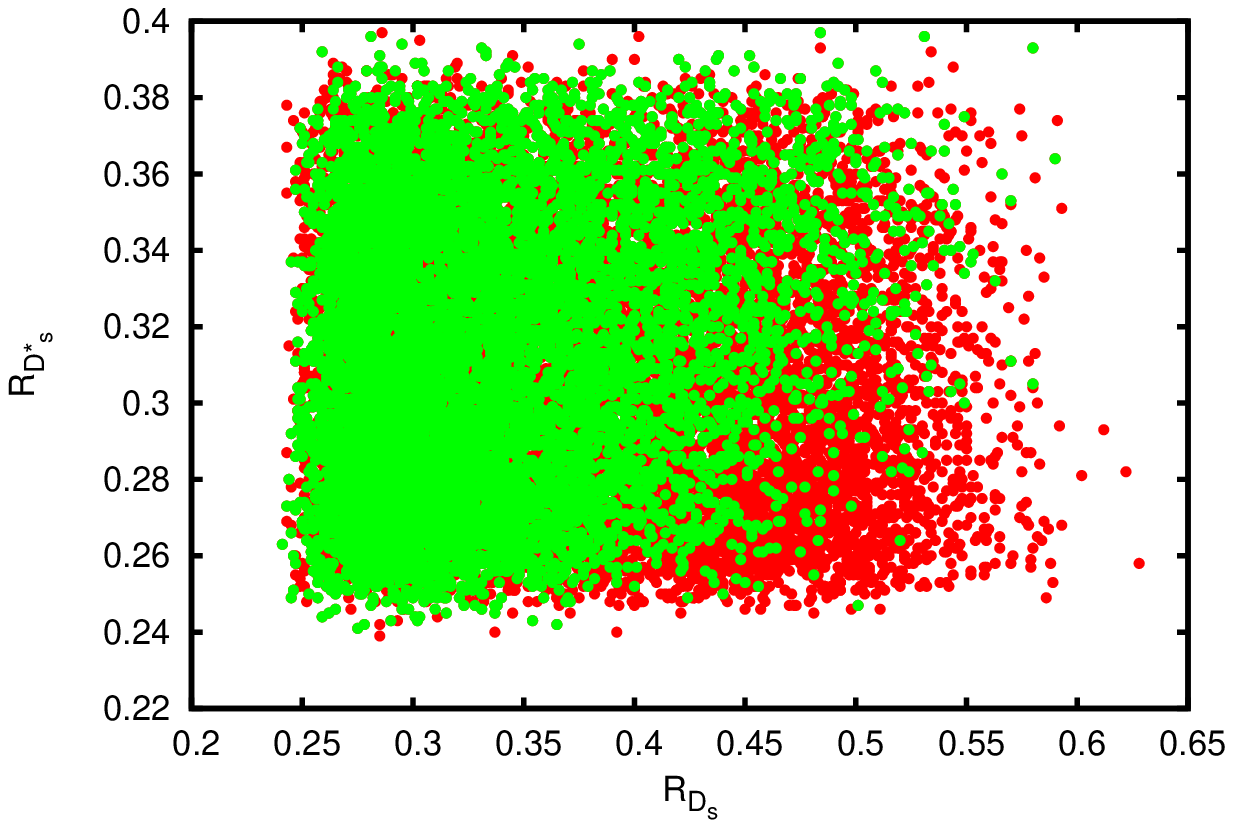}
\includegraphics[width=6cm,height=4cm]{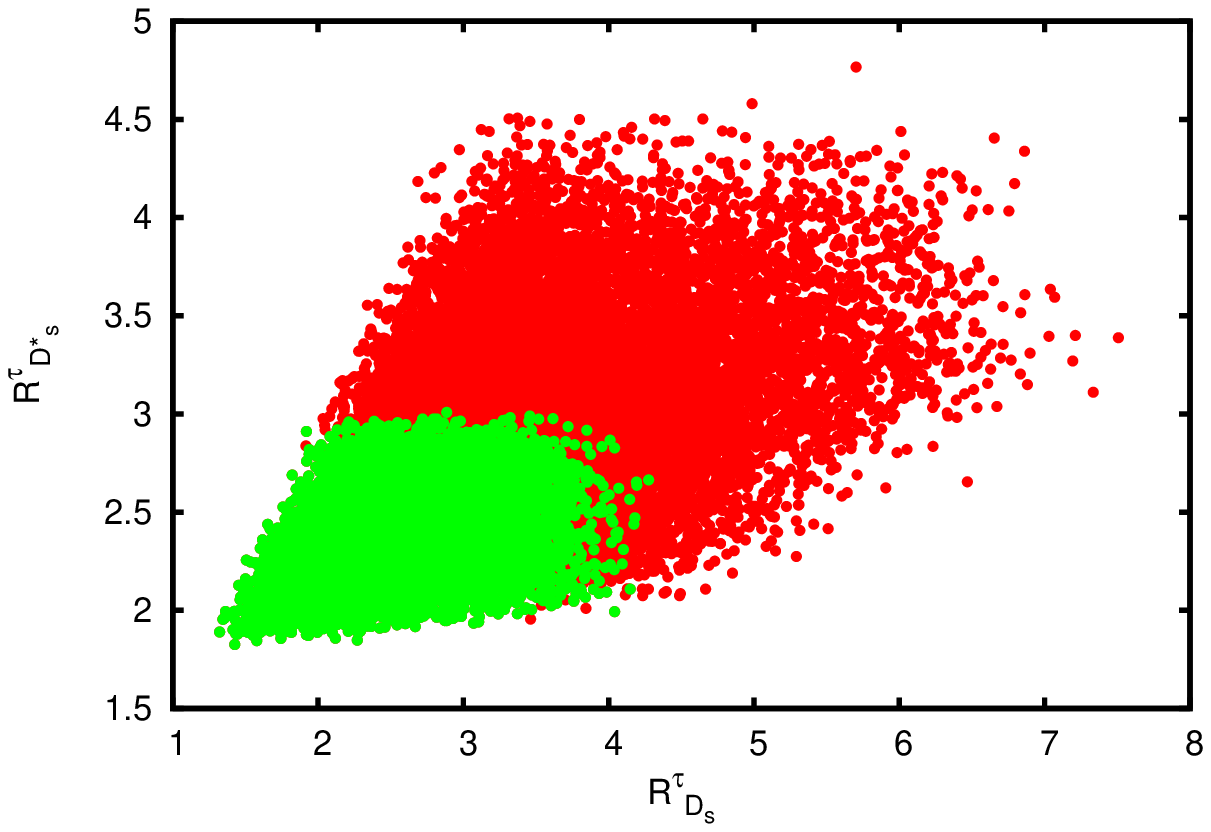}
\includegraphics[width=6cm,height=4cm]{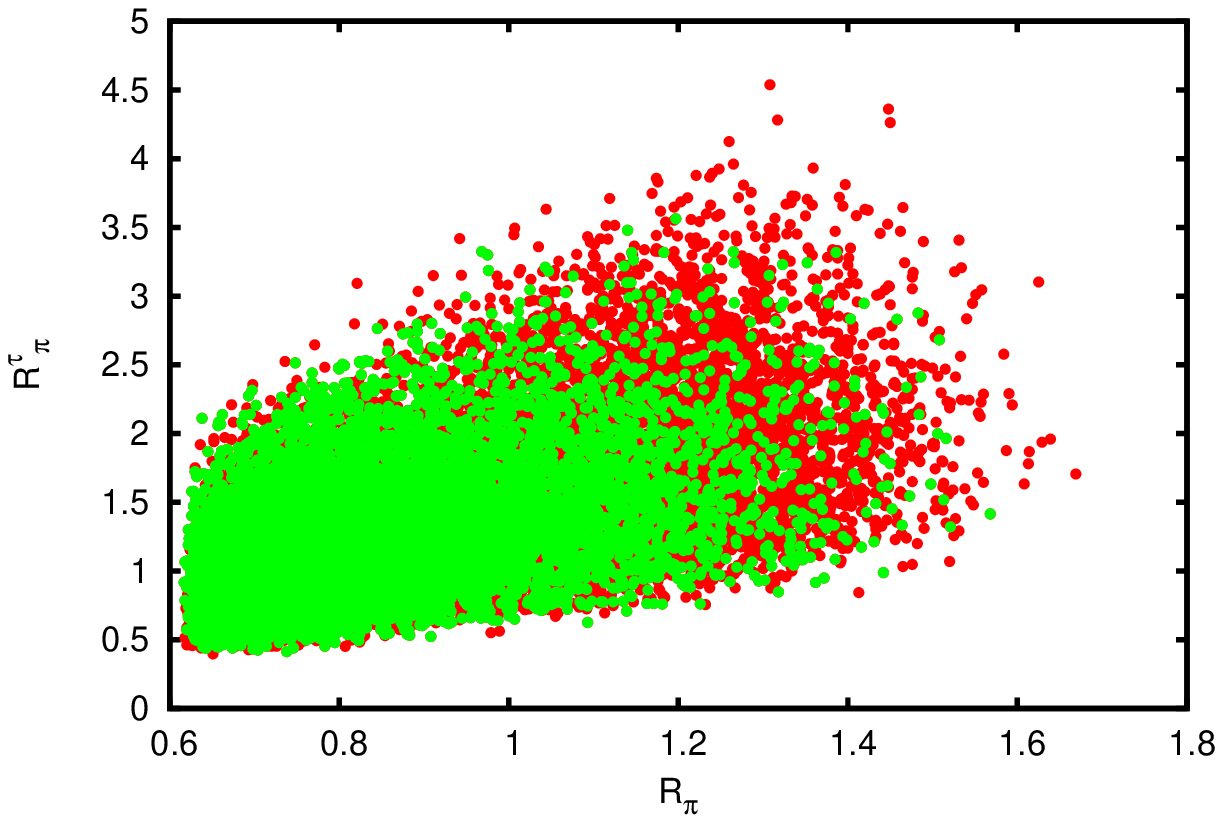}
\end{center}
\caption{\footnotesize{Allowed ranges in various observables with $\widetilde{V}_L$ and $\widetilde{V}_R$ type NP couplings once BABAR 
$2\sigma$ experimental constraint is imposed. We show in dark~(red) the allowed ranges once $2\sigma$ constraints from
$R_D$, $R_{D^{\ast}}$, and $\mathcal B(B \to \tau\nu)$ are imposed. Similarly, the allowed ranges are shown in light~(green)
once additional $2\sigma$ constraints from $R_D^{\tau}$ and $R_{D^{\ast}}^{\tau}$ are imposed.}} 
\label{obs_bab_np3}
\end{figure}

In the third scenario, we study the impact of new vector couplings $\widetilde{V}_L$ and 
$\widetilde{V}_R$, associated with right handed neutrinos, on various observables.
We first restrict the NP parameter space by imposing $2\sigma$ experimental constraints coming from the BABAR measured
values of the ratio of branching ratios $R_D$, $R_{D^{\ast}}$, and $\mathcal B(B \to \tau\nu)$. We also impose $2\sigma$ 
constraints coming from
the values of $R_D^{\tau}$ and $R_{D^{\ast}}^{\tau}$ that are estimated using the BABAR measured values of $R_D$ and $R_{D^{\ast}}$, 
and $\mathcal B(B \to \tau\nu)$. The NP effect coming from $\widetilde{V}_L$ and
$\widetilde{V}_R$ on various observables are shown in Fig.~\ref{obs_bab_np3}. 
We report the ranges in each observable in Table.~\ref{tab6}.
\begin{table}[htdp]
\vspace*{-0.1cm}
\begin{center}
\begin{tabular}{||c|c|c||c|c|c||}
\hline
Observable & Column I  & Column II &Observable & Column I& Column II\\
\hline
\hline
$R_D^{\tau}(\times 10^3)$ &$(1.521, 7.346)$ &$(1.521, 4.080)$ &$R_{\pi}^{\tau}$ &$(0.398, 4.538)$ &$(0.414, 3.563)$ \\[0.2cm]
\hline
$R_{D^{\ast}}^{\tau}(\times 10^3)$ &$(1.889, 4.841)$ &$(1.889, 2.978)$ &$R_{\pi}$ &$(0.617, 1.669)$ &$(0.617, 1.568)$ \\[0.2cm]
\hline
$R_{D_s}^{\tau}(\times 10^3)$ &$(1.323, 7.508)$ &$(1.323, 4.275)$ &$R_{D_s}$ &$(0.241, 0.628)$ &$(0.241, 0.590)$ \\[0.2cm]
\hline
$R_{{D^{\ast}_s}}^{\tau}(\times 10^3)$ &$(1.826, 4.767)$ &$(1.826, 3.008)$ &$R_{D^{\ast}_s}$ &$(0.239, 0.397)$ &$(0.241, 0.397)$ \\[0.2cm]
\hline
\hline
\end{tabular}
\end{center}
\caption{\footnotesize{Allowed ranges in various observables with $(\widetilde{V}_L,\,\widetilde{V}_R)$ NP couplings. 
The ranges reported in Column I represent 
the allowed values of each observable once constraints coming from BABAR
measured values of $R_D$, $R_{D^{\ast}}$, and $\mathcal B(B \to \tau\nu)$ are imposed. Column II represents the allowed range in
each observable once additional $2\sigma$ constraints from $R_D^{\tau}$
and $R_{D^{\ast}}^{\tau}$ are imposed.}}
\label{tab6}
\end{table}
We see significant deviation of all the observables from the SM prediction similar to the first scenario. We observe that the 
ranges in $R_D^{\tau}$, $R_{D^{\ast}}^{\tau}$, 
$R_{D_s}^{\tau}$, $R_{D^{\ast}_s}^{\tau}$, and $R_{\pi}^{\tau}$ do reduce once the additional $2\sigma$ constraints coming from 
$R_D^{\tau}$ and $R_{D^{\ast}}^{\tau}$ are imposed. However, we see no or very little change in $R_{\pi}$, $R_{D_s}$, and
$R_{D^{\ast}_s}$. Again, if only $\widetilde{V}_L$ type NP couplings were present, then $\widetilde{G}_V = \widetilde{G}_A$ 
and the NP effect will cancel in $R^\tau_{D_{(s)}^{(\ast)}}$ and $R_{\pi}^{\tau}$.

In the fourth scenario, we vary $\widetilde{S}_L$ and $\widetilde{S}_R$, new scalar couplings associated with right handed
neutrinos, while keeping others to zero. We find that only one set of $\widetilde{S}_L,\,\widetilde{S}_R$ namely 
$\widetilde{S}_L =  0.467$ and $\widetilde{S}_R = 0.003$ satisfy  
the $2\sigma$ experimental constraint coming from the BABAR measured values of the 
ratio of branching ratios $R_D$, $R_{D^{\ast}}$, and $\mathcal B(B \to \tau\nu)$. Corresponding values of all the observables are
tabulated in Table.~\ref{tab7}. 
\begin{table}[htdp]
\vspace*{-0.1cm}
\begin{center}
\begin{tabular}{||c|c|c||c|c|c||}
\hline
Observable & Column I  & Column II &Observable & Column I& Column II\\
\hline
\hline
$R_D^{\tau}(\times 10^3)$ &$1.597$ &$1.597$ &$R_{\pi}^{\tau}$ &$0.392$ & $0.392$\\[0.2cm]
\hline
$R_{D^{\ast}}^{\tau}(\times 10^3)$ &$1.052$ &$1.052$ &$R_{\pi}$ &$1.060$ &$1.060$ \\[0.2cm]
\hline
$R_{D_s}^{\tau}(\times 10^3)$ &$1.305$ &$1.305$ &$R_{D_s}$ &$0.338$ &$0.338$ \\[0.2cm]
\hline
$R_{{D^{\ast}_s}}^{\tau}(\times 10^3)$ &$0.943$ &$0.943$ &$R_{D^{\ast}_s}$ &$0.244$ &$0.244$ \\[0.2cm]
\hline
\hline
\end{tabular}
\end{center}
\caption{\footnotesize{Allowed values of various observables with $\widetilde{S}_L$ and $\widetilde{S}_R$ type NP couplings 
once the BABAR $2\sigma$ experimental
constraint is imposed. Allowed values~(Column I) obtained by imposing $2\sigma$ constraint coming from $R_D$, $R_{D^{\ast}}$,
and $\mathcal B(B \to \tau\nu)$ overlaps with the allowed values~(Column II) once additional $2\sigma$
constraint from $R_D^{\tau}$ and $R_{D^{\ast}}^{\tau}$ is imposed.}}
\label{tab7}
\end{table}
Again significant deviation from the SM expectation is observed for all the observables.  Imposing the additional 
$2\sigma$ constraints coming from the new observables $R_D^{\tau}$ and $R_{D^{\ast}}^{\tau}$ do not seem to affect the observables
in this scenario. 

It is observed that all the NP scenarios can accommodate the existing data on $b \to (u,\,c)\tau\nu$ decays. However,
for $S_L$ and $S_R$ type NP couplings there are very few points that are compatible with
the $2\sigma$ constraints coming from BABAR measurements. Similarly, for $\widetilde{S}_L$ and $\widetilde{S}_R$
type NP couplings there is only one set of points that satisfy the BABAR $2\sigma$ constraints. It is worth mentioning that
more precise data on $R_D^{\tau}$ and $R_{D^{\ast}}^{\tau}$ will be crucial in distinguishing various NP structures.
\subsection{BELLE constraint}
\label{belle}
\begin{figure}
\begin{center}
\includegraphics[width=6cm,height=4cm]{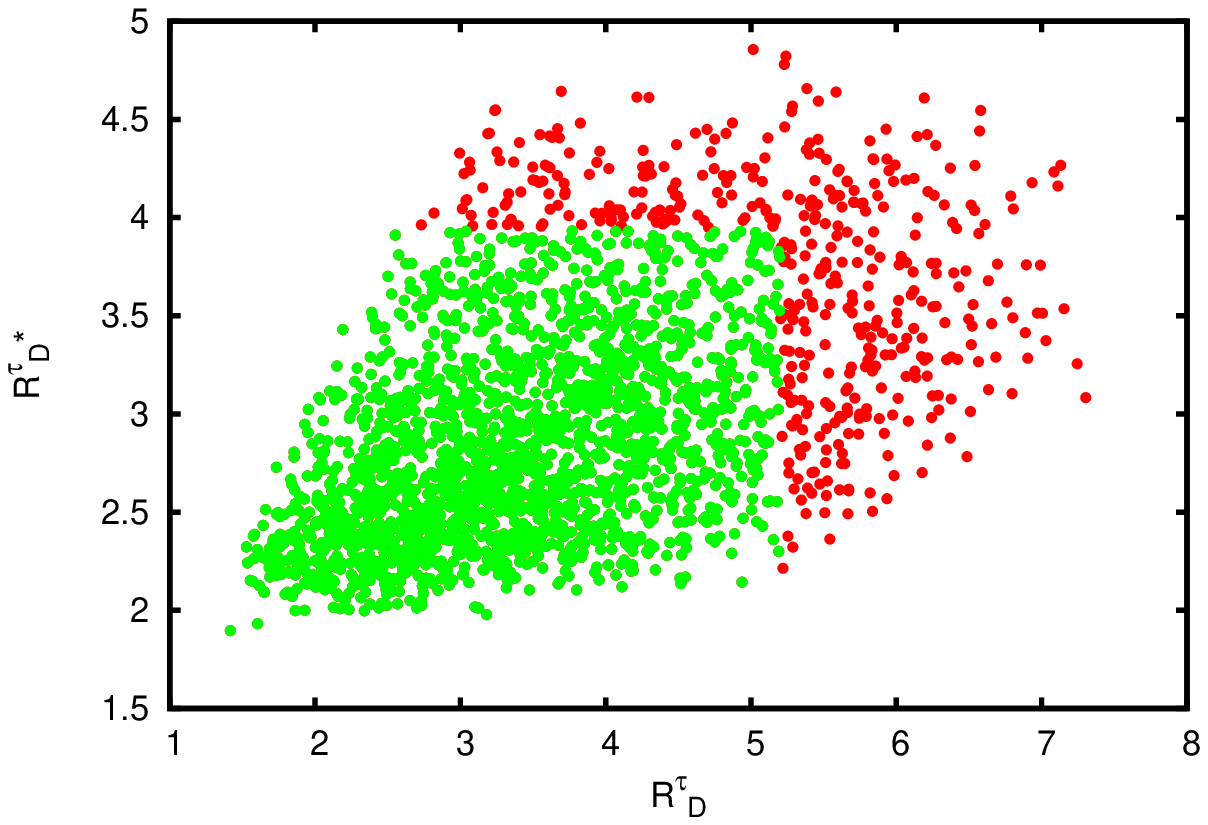}
\includegraphics[width=6cm,height=4cm]{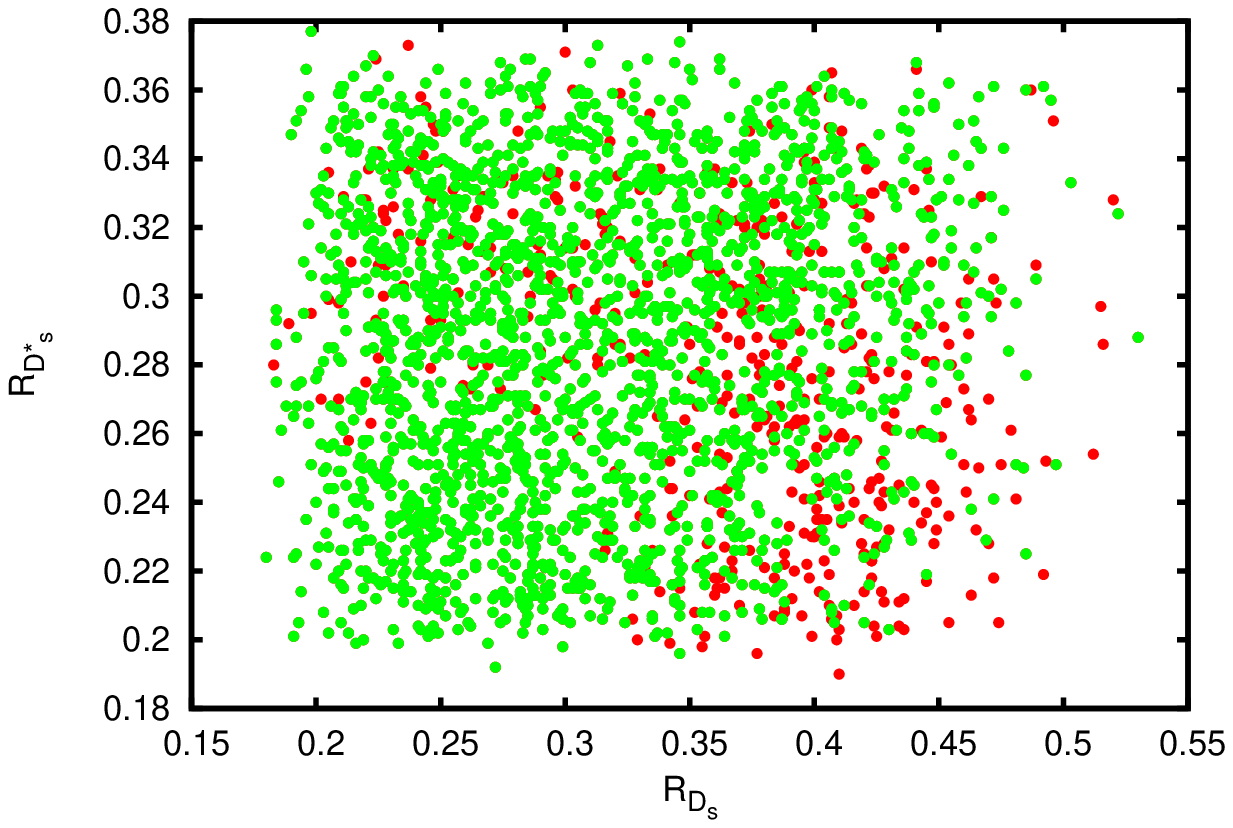}
\includegraphics[width=6cm,height=4cm]{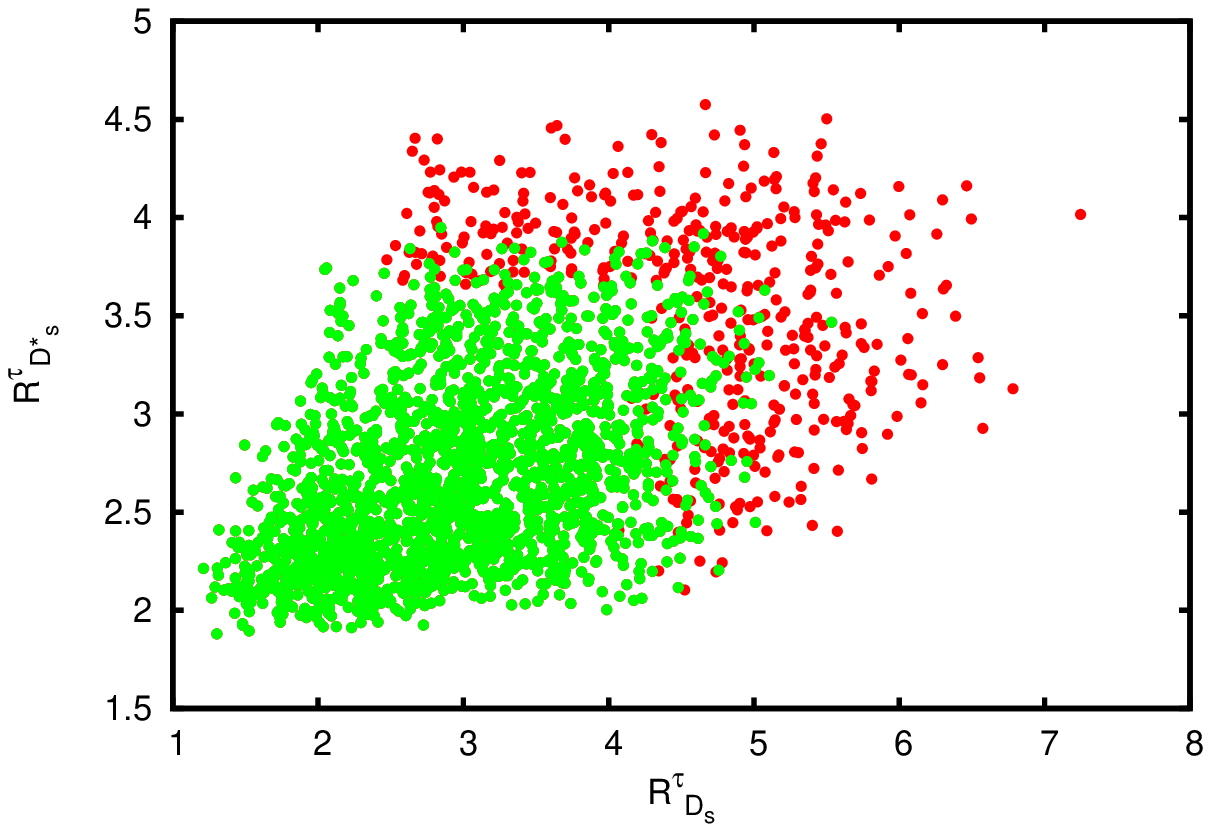}
\includegraphics[width=6cm,height=4cm]{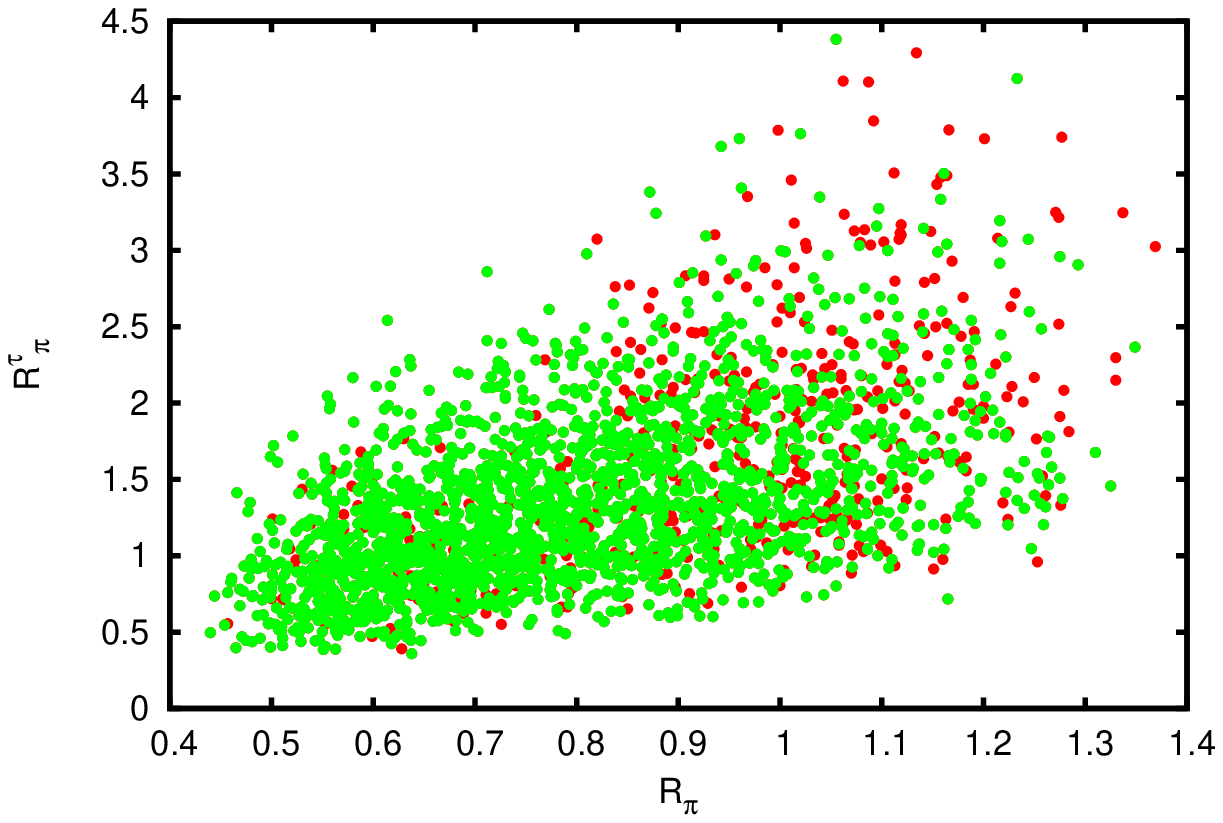}
\end{center}
\caption{\footnotesize{Allowed ranges in various observables with $V_L$ and $V_R$ type NP couplings once BELLE $2\sigma$ 
experimental constraint is imposed. Dark~(red) regions represent the allowed range obtained by imposing $2\sigma$
constraints coming from BELLE measured values of $R_D$, $R_{D^{\ast}}$, and $\mathcal B(B \to \tau\nu)$, whereas, 
the light~(green) regions represent the allowed range once $2\sigma$ additional constraints from $R_D^{\tau}$ and
$R_{D^{\ast}}^{\tau}$ are imposed.}} 
\label{obs_bel_np1}
\end{figure}

\begin{figure}
\begin{center}
\includegraphics[width=6cm,height=4cm]{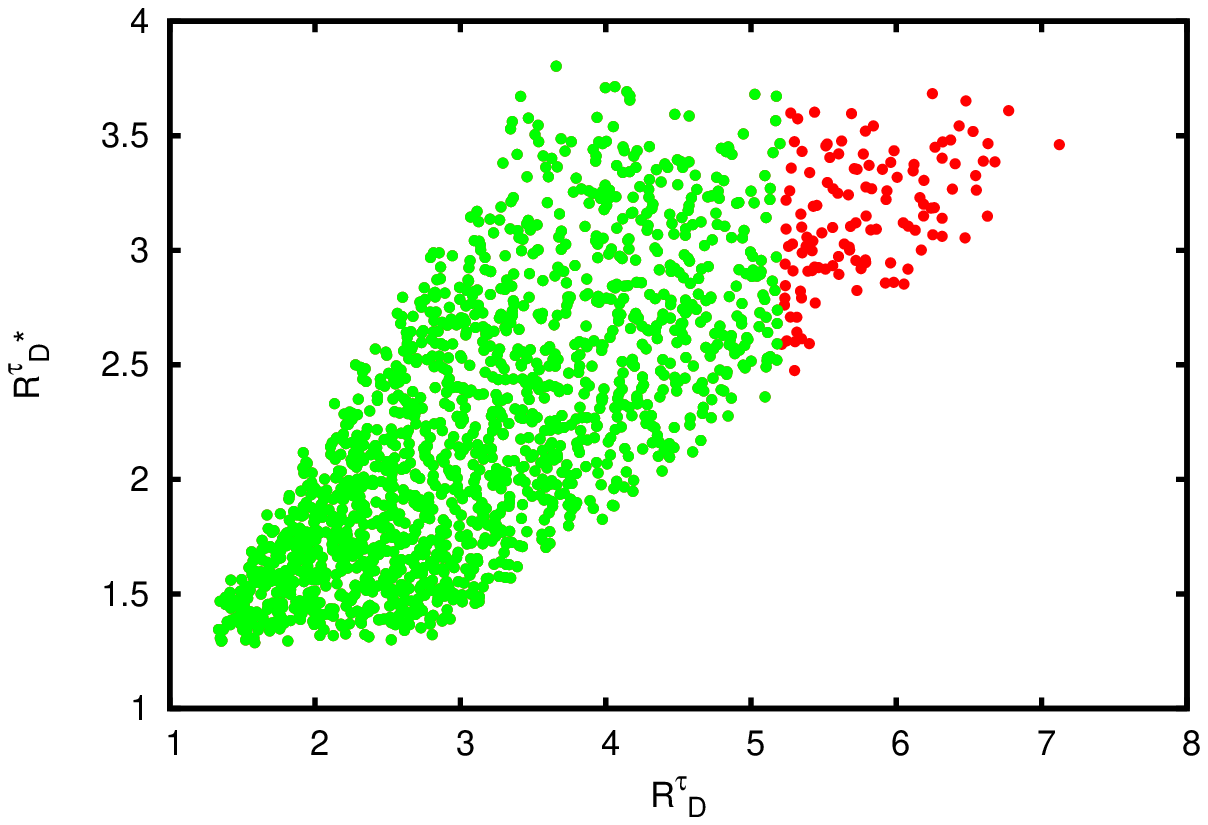}
\includegraphics[width=6cm,height=4cm]{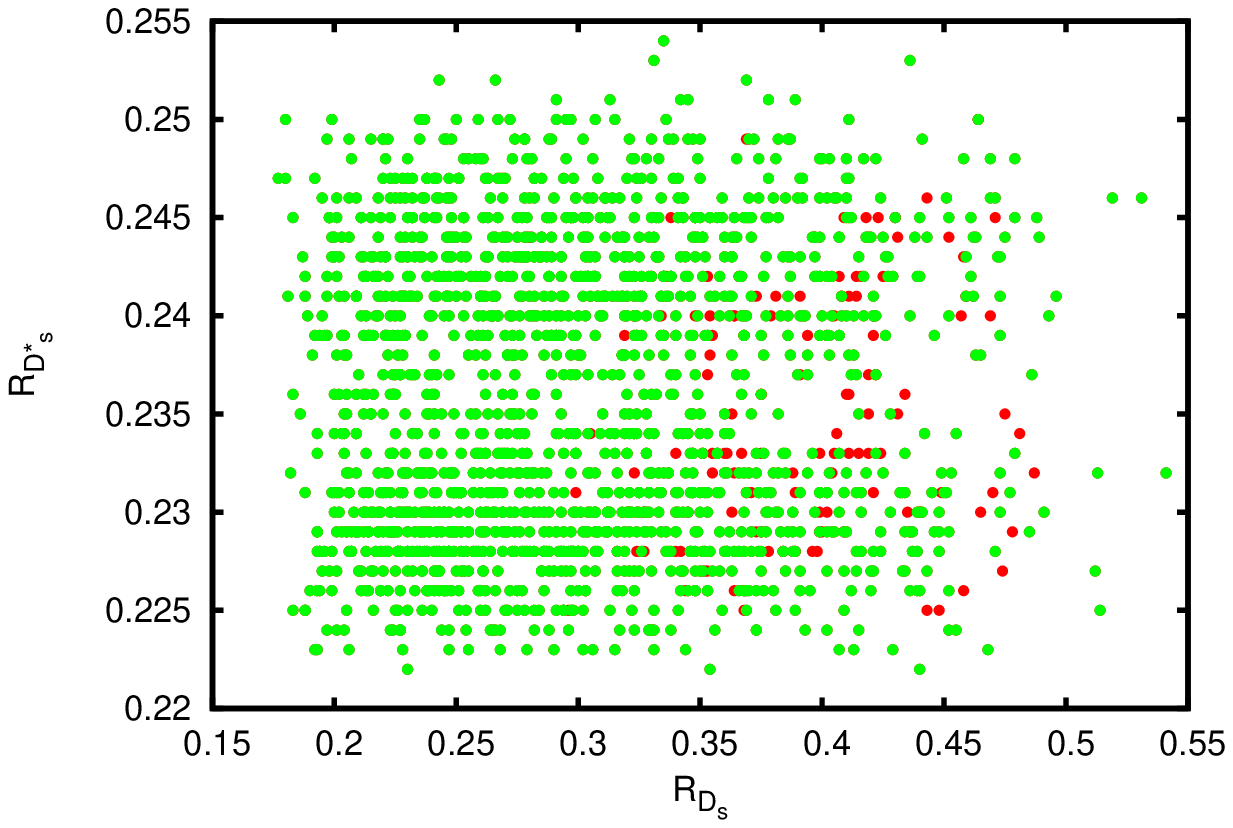}
\includegraphics[width=6cm,height=4cm]{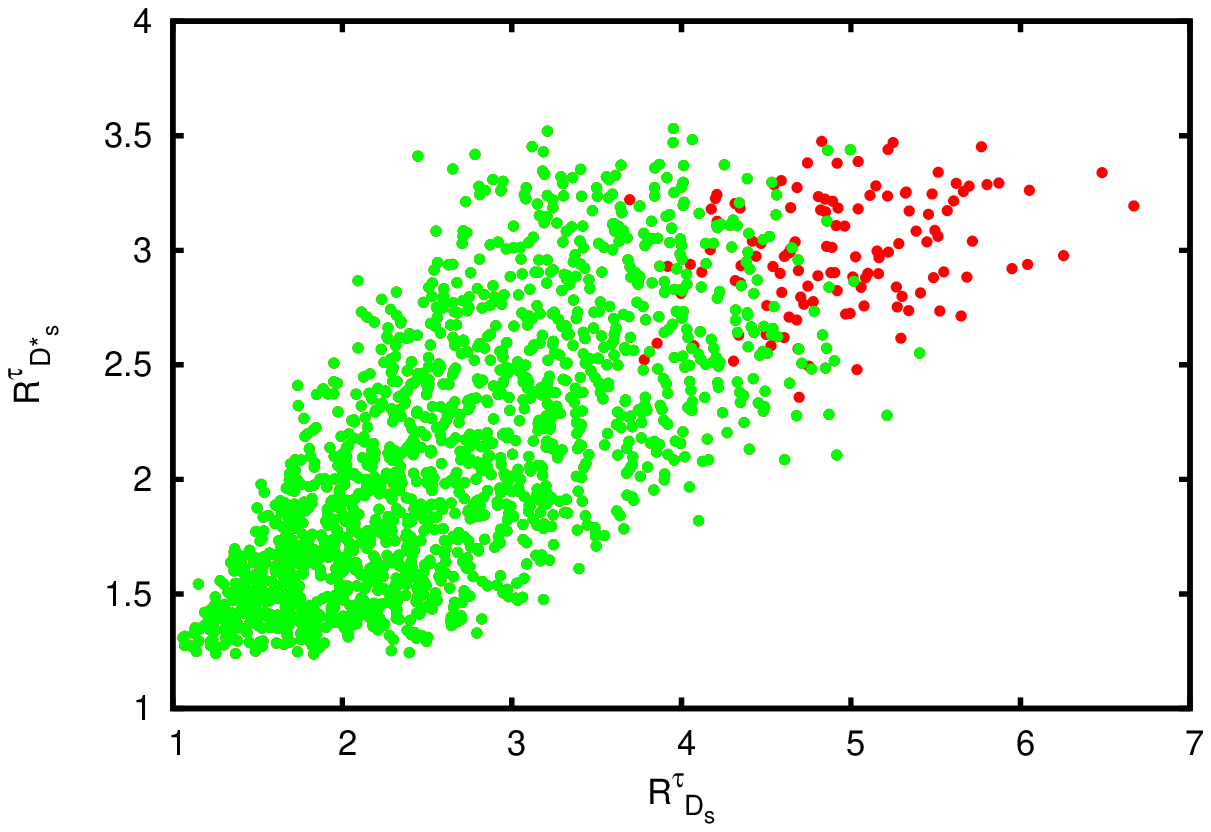}
\includegraphics[width=6cm,height=4cm]{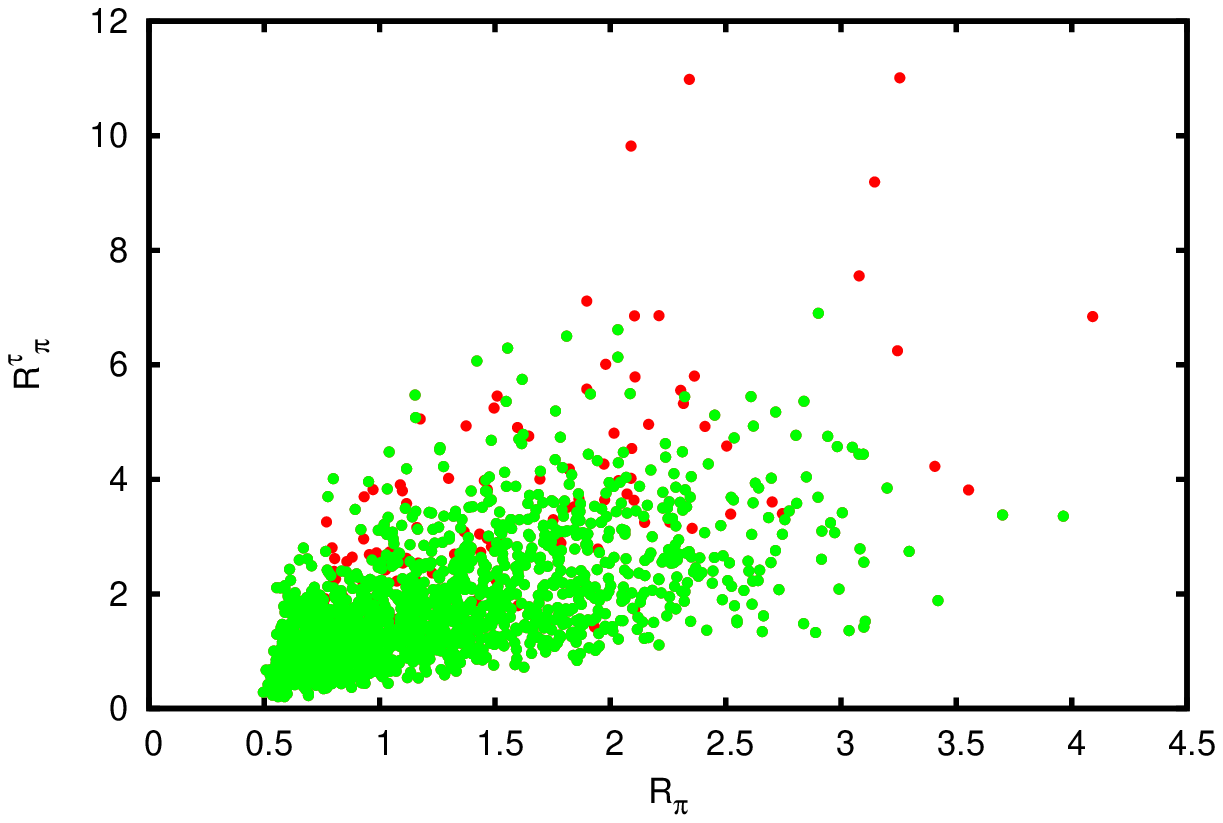}
\end{center}
\caption{\footnotesize{Allowed ranges in various observables with $S_L$ and $S_R$ type NP couplings once BELLE $2\sigma$ 
experimental constraint is imposed. We show in dark~(red) the allowed range in each observable once $2\sigma$
constraints coming from BELLE measured values of $R_D$, $R_{D^{\ast}}$, and $\mathcal B(B \to \tau\nu)$ are imposed.
Again, we show in light~(green) the allowed range once additional $2\sigma$
constraints from $R_D^{\tau}$ and $R_{D^{\ast}}^{\tau}$ are imposed.}} 
\label{obs_bel_np2}
\end{figure}

\begin{figure}
\begin{center}
\includegraphics[width=6cm,height=4cm]{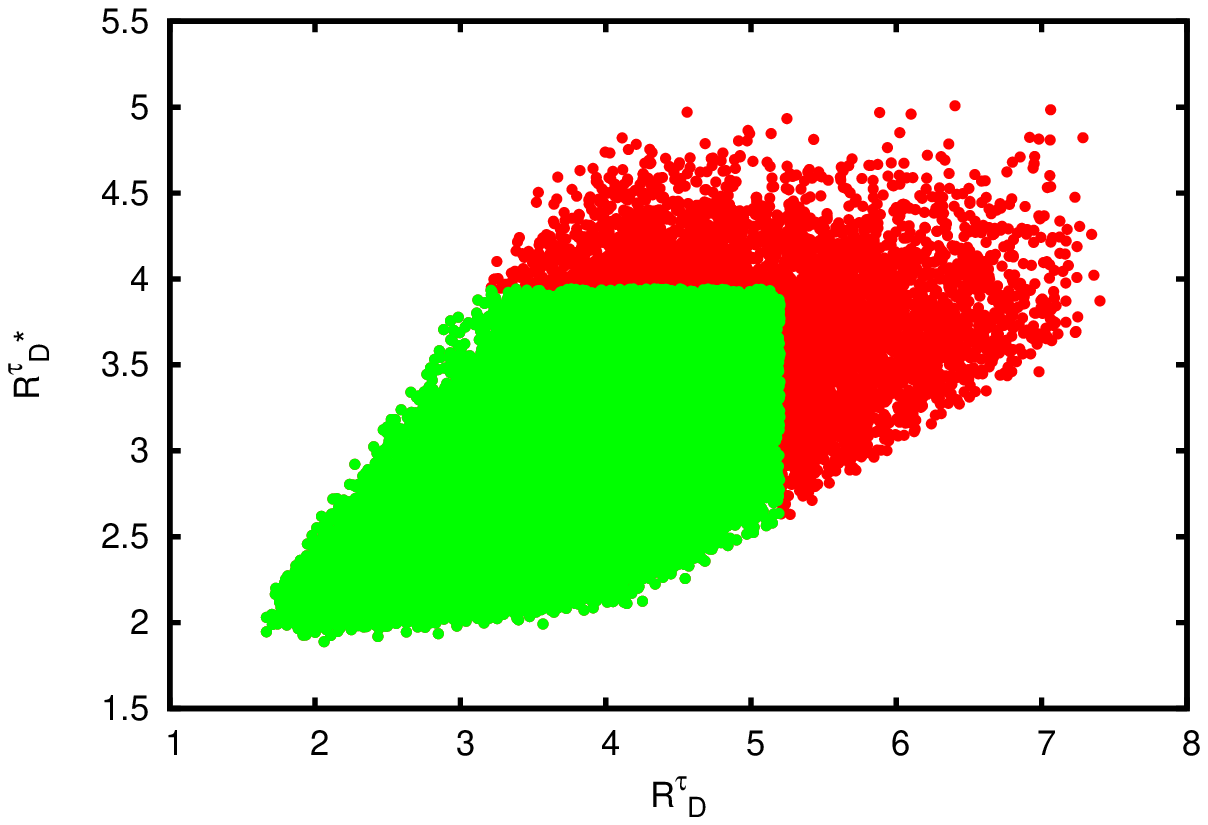}
\includegraphics[width=6cm,height=4cm]{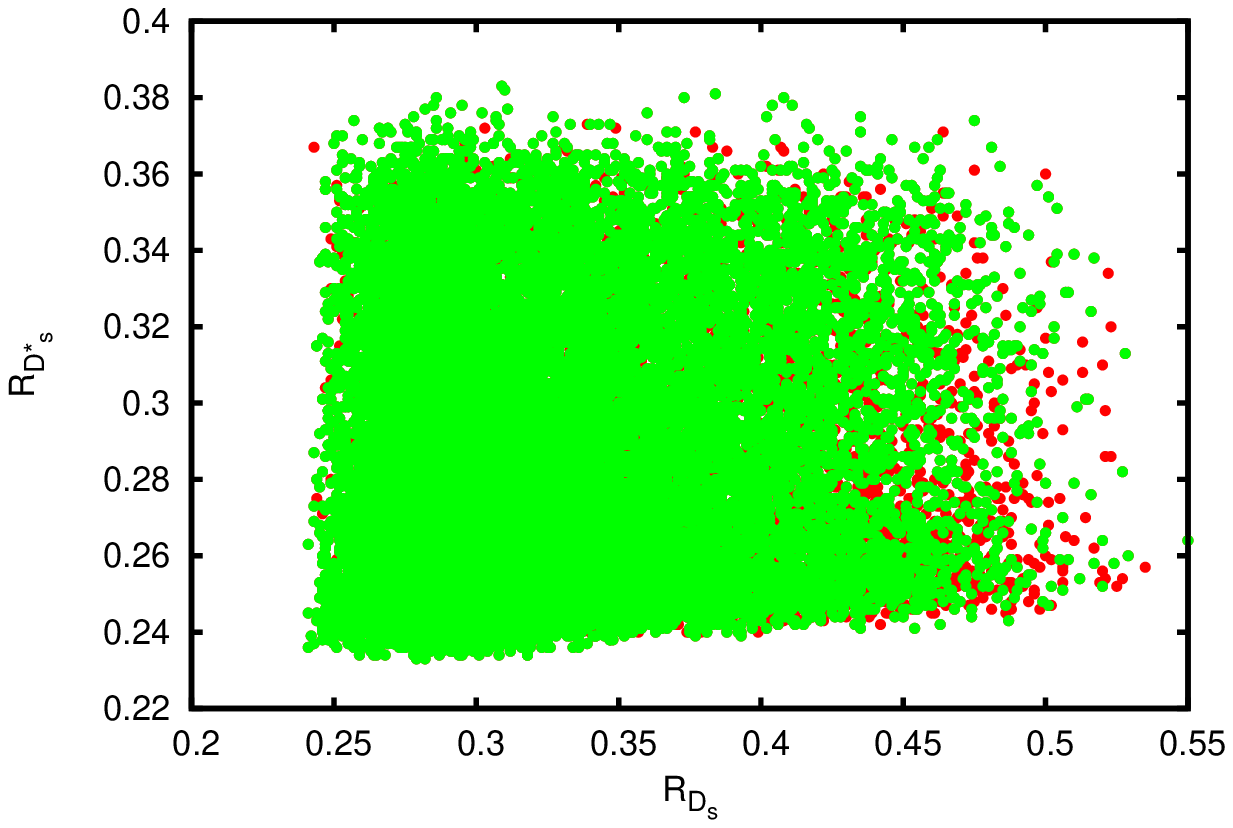}
\includegraphics[width=6cm,height=4cm]{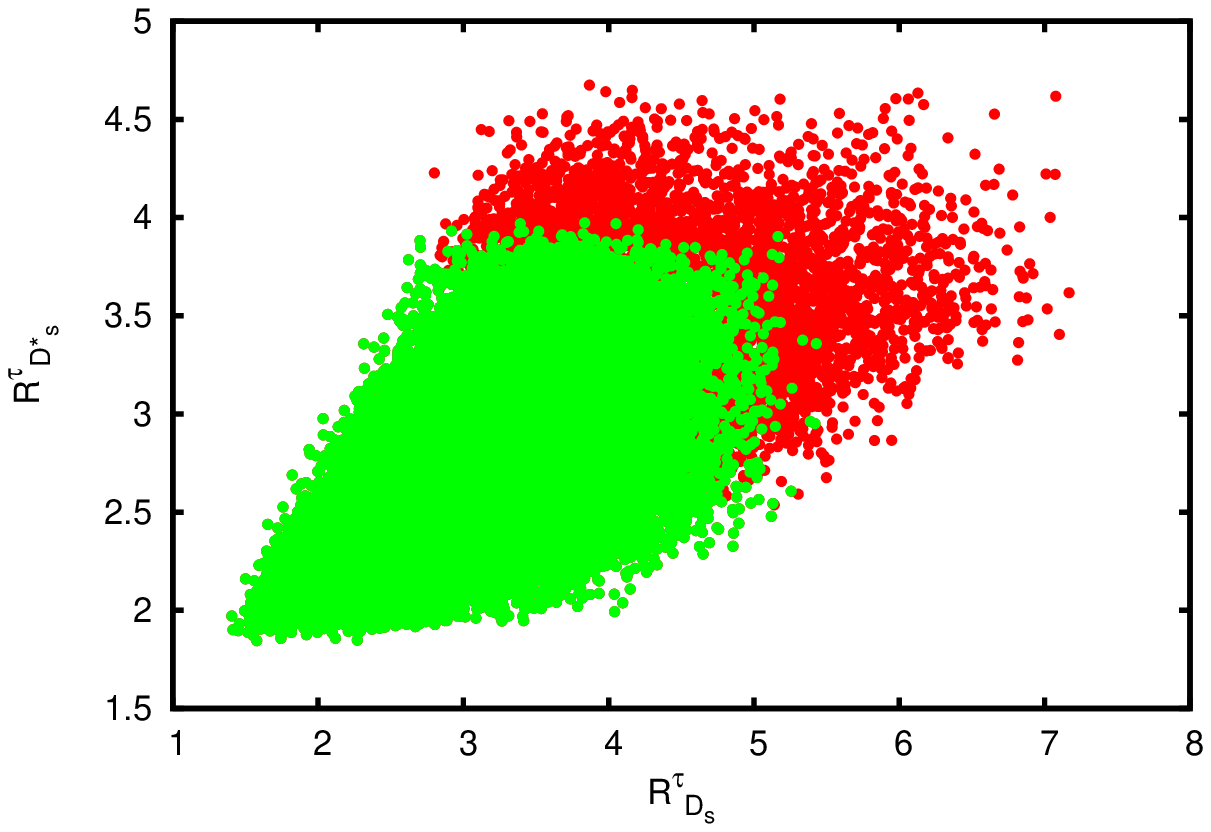}
\includegraphics[width=6cm,height=4cm]{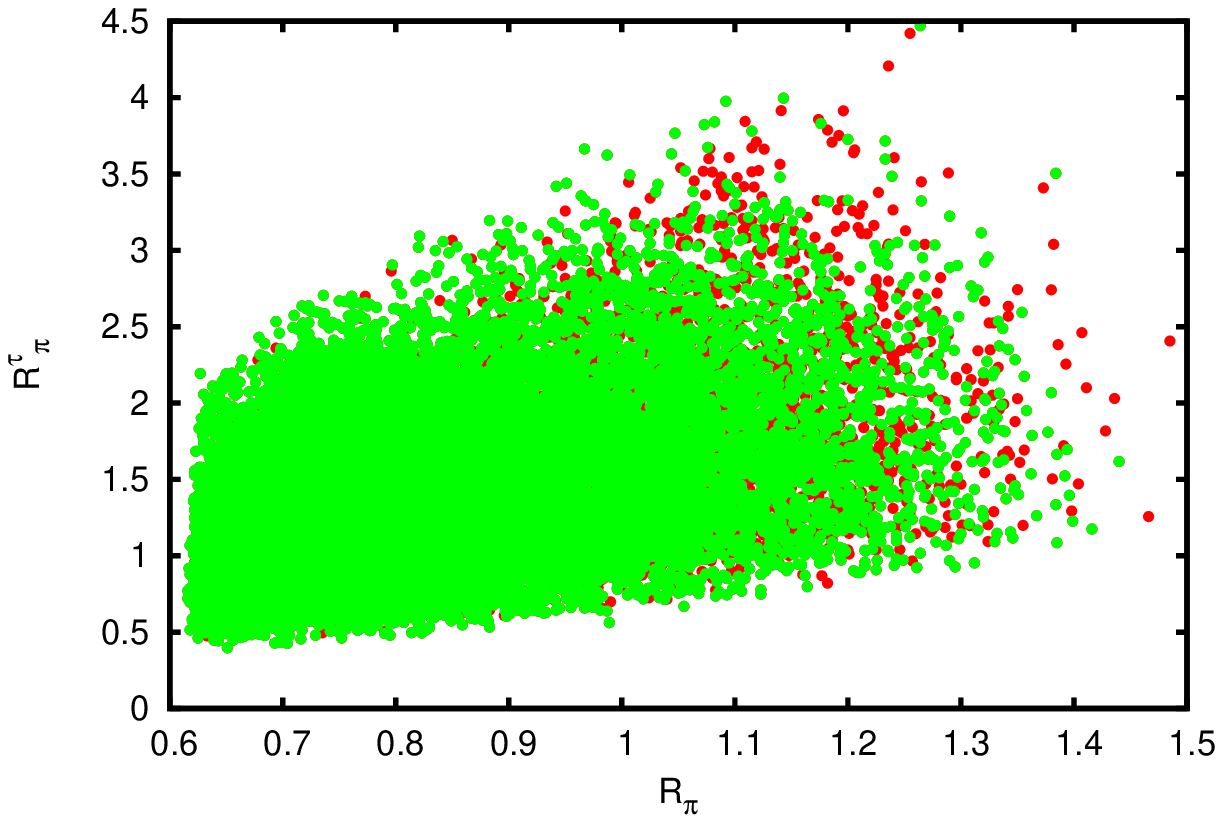}
\end{center}
\caption{\footnotesize{Allowed ranges in various observables with $\widetilde{V}_L$ and $\widetilde{V}_R$ type NP couplings 
once BELLE $2\sigma$ experimental constraint is imposed. Dark~(red) regions represent the allowed range obtained by imposing $2\sigma$
constraints coming from BELLE measured values of $R_D$, $R_{D^{\ast}}$, and $\mathcal B(B \to \tau\nu)$, whereas, 
the light~(green) regions represent the allowed range once $2\sigma$ additional constraints from $R_D^{\tau}$ and
$R_{D^{\ast}}^{\tau}$ are imposed.}} 
\label{obs_bel_np3}
\end{figure}

\begin{figure}
\begin{center}
\includegraphics[width=6cm,height=4cm]{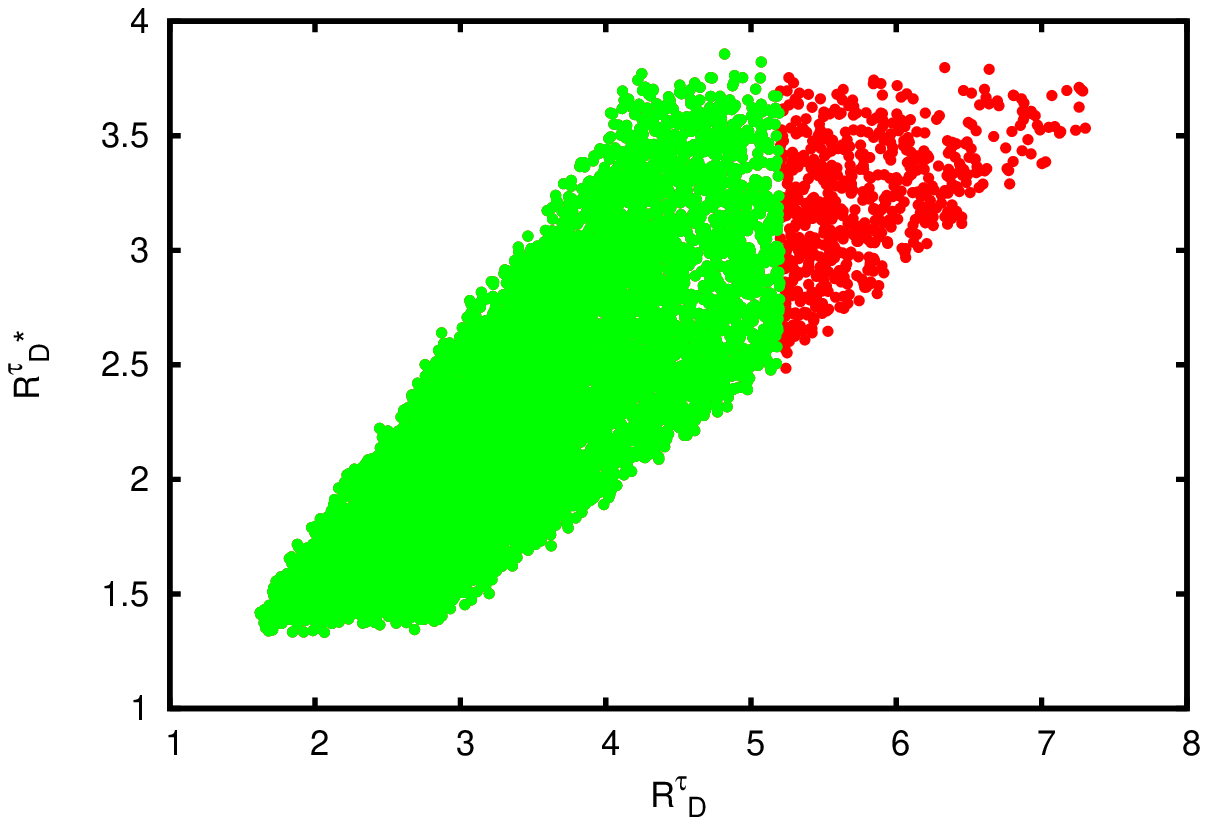}
\includegraphics[width=6cm,height=4cm]{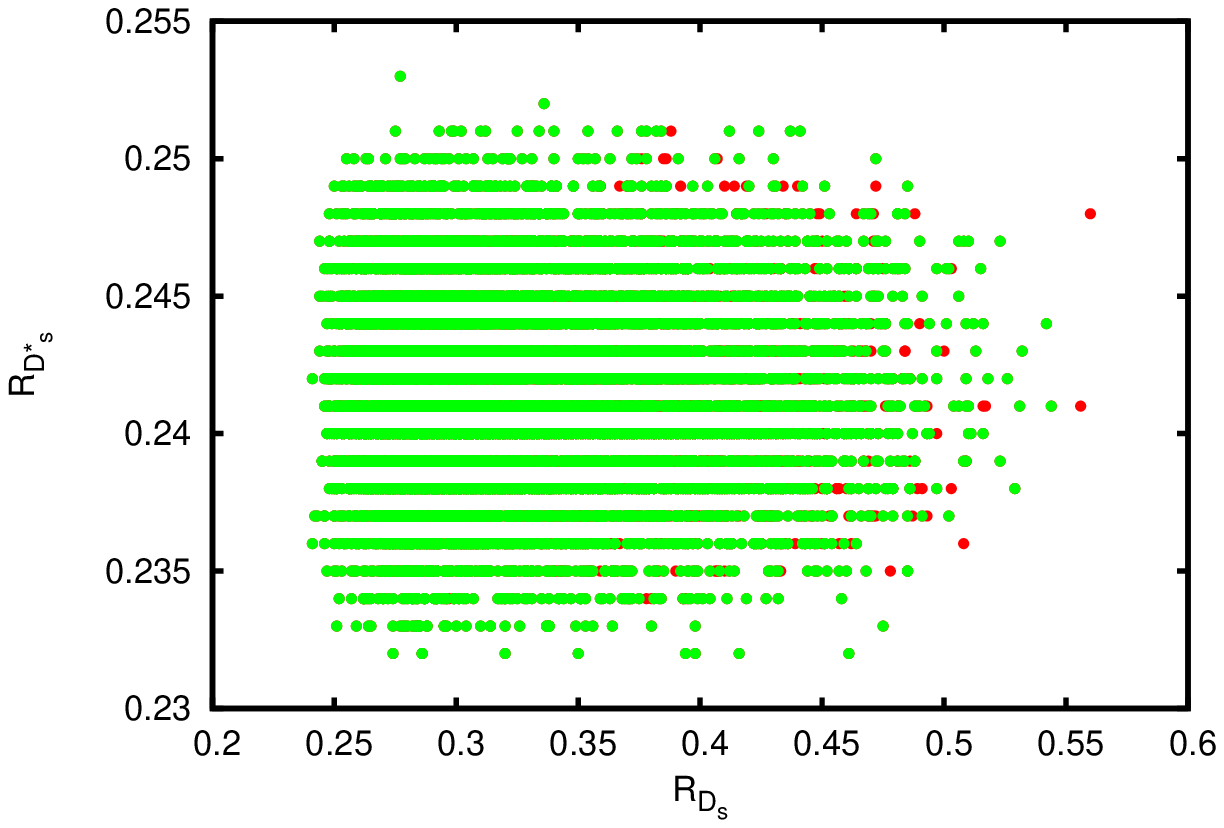}
\includegraphics[width=6cm,height=4cm]{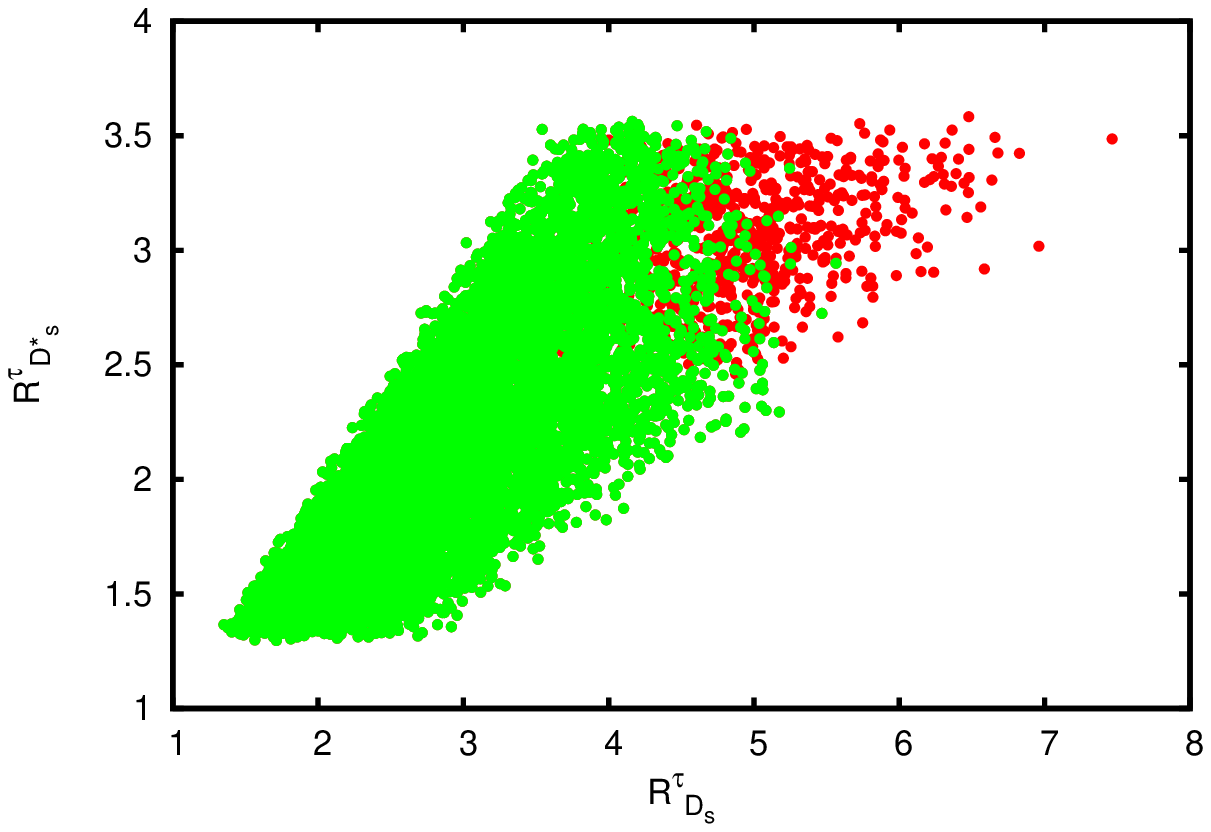}
\includegraphics[width=6cm,height=4cm]{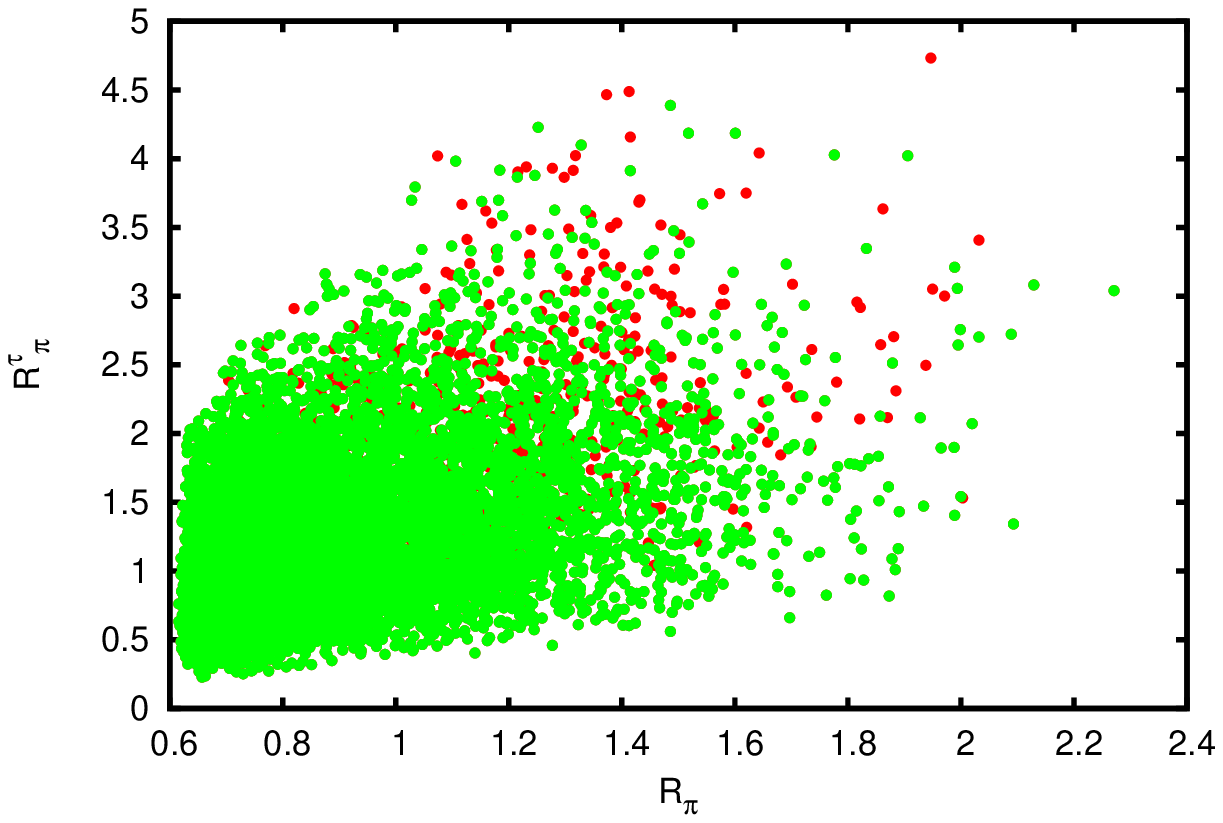}
\end{center}
\caption{\footnotesize{Allowed ranges in various observables with $\widetilde{S}_L$ and $\widetilde{S}_R$ type NP couplings once 
BELLE $2\sigma$ experimental constraint is imposed. We show in dark~(red) the allowed range in each observable once $2\sigma$
constraints coming from BELLE measured values of $R_D$, $R_{D^{\ast}}$, and $\mathcal B(B \to \tau\nu)$ are imposed.
Similarly, the allowed ranges in each observable are shown in light~(green) once additional $2\sigma$
constraints from $R_D^{\tau}$ and $R_{D^{\ast}}^{\tau}$ are imposed.}} 
\label{obs_bel_np4}
\end{figure}

Now we wish to find the effect  of $(V_L,\,V_R)$, $(S_L,\,S_R)$, $(\widetilde{V}_L,\,\widetilde{V}_R)$, and 
$(\widetilde{S}_L,\,\widetilde{S}_R)$ type NP couplings on all the observables using experimental constraint 
coming from the BELLE measurement. We consider four
different NP scenarios similar to BABAR analysis in section.~\ref{babar}. Similar to BABAR analysis in section.~\ref{babar}, 
we first impose $2\sigma$ constraints coming from the BELLE measured values of the ratio of branching ratios $R_D$,
$R_{D^{\ast}}$, and $\mathcal B(B \to \tau\nu)$ to explore various NP scenarios. We again impose $2\sigma$ constraints coming from 
$R_D^{\tau}$ and $R_{D^{\ast}}^{\tau}$ that are estimated using the BELLE measured values of $R_D$, 
$R_{D^{\ast}}$, and $\mathcal B(B \to \tau\nu)$ to see whether it is possible to constrain the NP parameter space even further.
Effect of NP on each observable under various scenarios are shown in Fig.~\ref{obs_bel_np1}, Fig.~\ref{obs_bel_np2}, 
Fig.~\ref{obs_bel_np3}, and Fig.~\ref{obs_bel_np4}. 
\begin{table}[htdp]
\vspace*{-0.1cm}
\begin{center}
\begin{tabular}{||c|c|c||c|c|c||}
\hline
Observable & Column I  & Column II &Observable & Column I& Column II\\
\hline
\hline
$R_D^{\tau}(\times 10^3)$ &$(1.417, 7.304)$ &$(1.417, 5.198)$ &$R_{\pi}^{\tau}$ &$(0.359, 4.382)$ &$(0.359, 4.382)$ \\[0.2cm]
\hline
$R_{D^{\ast}}^{\tau}(\times 10^3)$ &$(1.897, 4.856)$ &$(1.897, 3.932)$ &$R_{\pi}$ &$(0.440, 1.369)$ &$(0.440, 1.349)$ \\[0.2cm]
\hline
$R_{D_s}^{\tau}(\times 10^3)$ &$(1.212, 7.247)$ &$(1.212, 5.534)$ &$R_{D_s}$ &$(0.180, 0.530)$ &$(0.180, 0.530)$ \\[0.2cm]
\hline
$R_{{D^{\ast}_s}}^{\tau}(\times 10^3)$ &$(1.880, 4.576)$ &$(1.880, 3.949)$ &$R_{D^{\ast}_s}$ &$(0.190, 0.377)$ &$(0.192, 0.377)$ \\[0.2cm]
\hline
\hline
\end{tabular}
\end{center}
\caption{\footnotesize{Allowed ranges in various observables with $(V_L,\,V_R)$ NP couplings. The ranges reported
in Column I represent the allowed values of each observable once constraints coming from BELLE
measured values of $R_D$, $R_{D^{\ast}}$, and $\mathcal B(B \to \tau\nu)$ are imposed, whereas, the ranges in Column II represent
the allowed values once additional $2\sigma$ constraints from $R_D^{\tau}$ and $R_{D^{\ast}}^{\tau}$ are imposed.}}
\label{tab8}
\end{table}
\begin{table}[htdp]
\vspace*{-0.1cm}
\begin{center}
\begin{tabular}{||c|c|c||c|c|c||}
\hline
Observable & Column I  & Column II &Observable & Column I& Column II\\
\hline
\hline
$R_D^{\tau}(\times 10^3)$ &$(1.336, 7.122)$ &$(1.336, 5.199)$ &$R_{\pi}^{\tau}$ &$(0.207, 11.012)$ &$(0.207, 6.901)$ \\[0.2cm]
\hline
$R_{D^{\ast}}^{\tau}(\times 10^3)$ &$(1.287, 3.804)$ &$(1.287, 3.804)$ &$R_{\pi}$ &$(0.496, 4.091)$ &$(0.496, 3.964)$ \\[0.2cm]
\hline
$R_{D_s}^{\tau}(\times 10^3)$ &$(1.060, 6.669)$ &$(1.060, 5.406)$ &$R_{D_s}$ &$(0.177, 0.541)$ &$(0.177, 0.541)$ \\[0.2cm]
\hline
$R_{{D^{\ast}_s}}^{\tau}(\times 10^3)$ &$(1.238, 3.532)$ &$(1.238, 3.532)$ &$R_{D^{\ast}_s}$ &$(0.222, 0.254)$ &$(0.222, 0.254)$ \\[0.2cm]
\hline
\hline
\end{tabular}
\end{center}
\caption{\footnotesize{Allowed ranges in various observables with $(S_L,\,S_R)$ NP couplings. The ranges reported
in Column I represent the allowed values of each observable once constraints coming from BELLE
measured values of $R_D$, $R_{D^{\ast}}$, and $\mathcal B(B \to \tau\nu)$ are imposed, whereas, the ranges in Column II represent
the allowed values once additional $2\sigma$ constraints from $R_D^{\tau}$ and $R_{D^{\ast}}^{\tau}$ are imposed.}}
\label{tab9}
\end{table}
\begin{table}[htdp]
\vspace*{-0.1cm}
\begin{center}
\begin{tabular}{||c|c|c||c|c|c||}
\hline
Observable & Column I  & Column II &Observable & Column I& Column II\\
\hline
\hline
$R_D^{\tau}(\times 10^3)$ &$(1.665, 7.402)$ &$(1.665, 5.200)$ &$R_{\pi}^{\tau}$ &$(0.398, 4.471)$ &$(0.398, 4.471)$ \\[0.2cm]
\hline
$R_{D^{\ast}}^{\tau}(\times 10^3)$ &$(1.889, 5.008)$ &$(1.889, 3.942)$ &$R_{\pi}$ &$(0.616, 1.485)$ &$(0.616, 1.440)$ \\[0.2cm]
\hline
$R_{D_s}^{\tau}(\times 10^3)$ &$(1.406, 7.168)$ &$(1.406, 5.429)$ &$R_{D_s}$ &$(0.241, 0.550)$ &$(0.241, 0.550)$ \\[0.2cm]
\hline
$R_{{D^{\ast}_s}}^{\tau}(\times 10^3)$ &$(1.846, 4.675)$ &$(1.846, 3.974)$ &$R_{D^{\ast}_s}$ &$(0.233, 0.383)$ &$(0.233, 0.383)$ \\[0.2cm]
\hline
\hline
\end{tabular}
\end{center}
\caption{\footnotesize{Allowed ranges in various observables with $(\widetilde{V}_L,\,\widetilde{V}_R)$ NP couplings. The ranges reported
in Column I represent the allowed values of each observable once constraints coming from BELLE
measured values of $R_D$, $R_{D^{\ast}}$, and $\mathcal B(B \to \tau\nu)$ are imposed, whereas, the ranges in Column II represent
the allowed values once additional $2\sigma$ constraints from $R_D^{\tau}$ and $R_{D^{\ast}}^{\tau}$ are imposed.}}
\label{tab10}
\end{table}
\begin{table}[htdp]
\vspace*{-0.1cm}
\begin{center}
\begin{tabular}{||c|c|c||c|c|c||}
\hline
Observable & Column I  & Column II &Observable & Column I& Column II\\
\hline
\hline
$R_D^{\tau}(\times 10^3)$ &$(1.622, 7.301)$ &$(1.622, 5.200)$ &$R_{\pi}^{\tau}$ &$(0.229, 4.732)$ &$(0.229, 4.387)$ \\[0.2cm]
\hline
$R_{D^{\ast}}^{\tau}(\times 10^3)$ &$(1.333, 3.857)$ &$(1.333, 3.857)$ &$R_{\pi}$ &$(0.616, 2.271)$ &$(0.616, 2.271)$ \\[0.2cm]
\hline
$R_{D_s}^{\tau}(\times 10^3)$ &$(1.352, 7.465)$ &$(1.352, 5.564)$ &$R_{D_s}$ &$(0.241, 0.560)$ &$(0.241, 0.544)$ \\[0.2cm]
\hline
$R_{{D^{\ast}_s}}^{\tau}(\times 10^3)$ &$(1.297, 3.583)$ &$(1.297, 3.563)$ &$R_{D^{\ast}_s}$ &$(0.232, 0.253)$ &$(0.232, 0.253)$ \\[0.2cm]
\hline
\hline
\end{tabular}
\end{center}
\caption{\footnotesize{Allowed ranges in various observables with $(\widetilde{S}_L,\,\widetilde{S}_R)$ NP couplings. The ranges reported
in Column I represent the allowed values of each observable once constraints coming from BELLE
measured values of $R_D$, $R_{D^{\ast}}$, and $\mathcal B(B \to \tau\nu)$ are imposed, whereas, the ranges in Column II represent
the allowed values once additional $2\sigma$ constraints from $R_D^{\tau}$ and $R_{D^{\ast}}^{\tau}$ are imposed.}}
\label{tab11}
\end{table}

The deviation from the SM expectation is found to be significant in all the four scenarios. 
The allowed ranges in each observable for each scenario are reported in Table.~\ref{tab8}, Table.~\ref{tab9}, Table.~\ref{tab10}, 
and Table.~\ref{tab11}. We see that for $(V_L,\,V_R)$ couplings, although, the allowed ranges of $R_D^{\tau}$, $R_{D^{\ast}}^{\tau}$, 
$R_{D_s}^{\tau}$, and $R_{D^{\ast}_s}^{\tau}$ do reduce, there is no or very little change in the allowed ranges of 
$R_{\pi}^{\tau}$, $R_{\pi}$, $R_{D_s}$, and 
$R_{D^{\ast}_s}$ once we impose $2\sigma$ constraints from $R_D^{\tau}$ and $R_{D^{\ast}}^{\tau}$. Similar results are 
observed for $(\widetilde{V}_L,\,\widetilde{V}_R)$ NP couplings as well. For $(S_L,\,S_R)$ type NP couplings, we find 
considerable reduction in the allowed ranges of $R_D^{\tau}$, $R_{D_s}^{\tau}$, and $R_{\pi}^{\tau}$, whereas, 
there is no or very little change in the allowed ranges of  $R_{D^{\ast}}^{\tau}$, $R_{D^{\ast}_s}^{\tau}$, $R_{D_s}$, 
$R_{D^{\ast}_s}$, and $R_{\pi}$ once additional $2\sigma$
constraints from the estimated values of $R_D^{\tau}$ and $R_{D^{\ast}}^{\tau}$ are imposed. Similar results are obtained
for $(\widetilde{S}_L,\,\widetilde{S}_R)$ NP couplings as well.

It is evident that all the four NP scenarios not only accommodate the existing data on $R_D$, $R_{D^{\ast}}$, and 
$\mathcal B(B \to \tau\nu)$, 
but also accommodate the newly estimated data on $R_D^{\tau}$ and $R_{D^{\ast}}^{\tau}$. Recent result from BELLE  
on $\mathcal B(B \to \pi\tau\nu) < 2.5 \times 10^{-4}$~\cite{Hamer:2015jsa} gives a upper limit on 
$R_{\pi}^{\tau} < 2.62$~\cite{Nandi:2016wlp}. It is worth 
mentioning that one can constrain the NP parameter space even further once more precise data on $\mathcal B(B \to \pi\tau\nu)$ 
is available.
Here too, more precise measurements are required to distinguish various NP structures.

\section{Conclusion}
\label{con}
Lepton flavor universality violation has been observed in various semileptonic $B$ meson decays. The measured values of 
$R_D$ and $R_{D^{\ast}}$ exceed the SM expectation by $1.9\sigma$ and $3.3\sigma$, respectively. HFAG reported 
the combined deviation from the SM prediction to be at the level of $4\sigma$. Similar tensions have been 
observed in $B \to (K,\,K^{\ast})\,l\,l$ and $B_s \to \phi\,l\,l$ decays mediated via $b \to s\,l\,l$ transition process as well. 
A lot of phenomenological 
studies have been performed in order to explain these discrepancies. Measurement of $B \to \tau\nu$ and 
$B \to (D,\,D^{\ast})\,\tau\nu$ decays suffer $\tau$ detection and identification systematics. To examine this possibility, very recently, 
in Ref.~\cite{Nandi:2016wlp}, the authors introduced two new observables namely $R_D^{\tau}$ and $R_{D^{\ast}}^{\tau}$ 
where the $\tau$ detection and identification systematics will largely cancel. The estimated values of  $R_D^{\tau}$ and $R_{D^{\ast}}^{\tau}$ 
are consistent with the SM prediction although there is discrepancy between the measured $R_D$ and 
$R_{D^{\ast}}$ with the SM prediction. This may occur for a class of NP which affect both 
$R_{D}$, $R_{D^{\ast}}$ and $B \to \tau\nu$ decays. In Ref.~\cite{Nandi:2016wlp}, the authors consider type II 
$2$HDM model to illustrate these points.

In this paper, we use an effective field theory in the presence of NP to explore various NP couplings 
in a model independent way. First, we consider the constraints coming from the 
measured values of $R_D$, $R_{D^{\ast}}$, and $\mathcal B(B \to \tau\nu)$ to see  
various NP effect on these new observables. Second, we see whether it is 
possible to constrain the NP parameter space even further by putting additional 
constraints coming from the estimated values of $R^{\tau}_D$ and $R^{\tau}_{D^{\ast}}$ 
since these ratios are consistent with the SM values. We study the effect of new 
physics couplings on various observables related to $B_s \to (D_s,\,D^{\ast}_s)\,\tau\nu$ and 
$B \to \pi\tau\nu$ decays as well.
The main results of our analysis are summarized below.

We first study the impact of NP couplings on various observables using $2\sigma$ constraints coming from BABAR measured values
of $R_D$, $R_{D^{\ast}}$, and 
$\mathcal B(B \to \tau\nu)$. We consider four different NP scenarios. We find significant 
deviation from the SM prediction in each observable for each scenario. We find that, although, each of the four NP scenarios can 
simultaneously explain all the existing data on $b \to u$ and $b \to c$ leptonic and semileptonic $B$ meson decays, 
there are very few points that are compatible within
the $2\sigma$ constraints coming from BABAR measurements for $(S_L,\,S_R)$ type NP couplings. Similarly, 
for $\widetilde{S}_L$ and $\widetilde{S}_R$ type NP couplings there is only one set of points that satisfy the BABAR 
$2\sigma$ constraints. 
Our second point was to see whether it is possible to constrain the NP parameter space even further by imposing constraints 
coming from the newly constructed observables $R_D^{\tau}$ and $R_{D^{\ast}}^{\tau}$ in a model independent way. We see that 
the additional constraint coming from the new observables $R_D^{\tau}$ and $R_{D^{\ast}}^{\tau}$ does not constrain $(S_L,\,S_R)$
and $(\widetilde{S}_L$,\, $\widetilde{S}_R)$ type NP parameter space. However, for $(V_L,\,V_R)$ and 
$(\widetilde{V}_L$,\, $\widetilde{V}_R)$ type NP couplings, 
the allowed ranges in $R_D^{\tau}$, $R_{D^{\ast}}^{\tau}$,
$R_{D_s}^{\tau}$, $R_{D^{\ast}_s}^{\tau}$, and $R_{\pi}^{\tau}$ are considerably reduced once the additional $2\sigma$ constraint 
from $R_D^{\tau}$ and $R_{D^{\ast}}^{\tau}$ are imposed.

We do the same analysis using the BELLE measured values. We first constrain the NP parameter space using $2\sigma$ constraints 
from BELLE measured values
of $R_D$, $R_{D^{\ast}}$, and $\mathcal B(B \to \tau\nu)$. 
The deviation from the SM expectation is found to be significant in all the four scenarios.
We find that for $(V_L,\,V_R)$ couplings, although, the allowed ranges in $R_D^{\tau}$, $R_{D^{\ast}}^{\tau}$,
$R_{D_s}^{\tau}$, and $R_{D^{\ast}_s}^{\tau}$ do reduce, there is no or very little change in $R_{\pi}^{\tau}$, $R_{\pi}$, $R_{D_s}$, and
$R_{D^{\ast}_s}$ allowed ranges once we impose $2\sigma$ constraints from $R_D^{\tau}$ and $R_{D^{\ast}}^{\tau}$. Similar results are
obtained for $(\widetilde{V}_L,\,\widetilde{V}_R)$ NP couplings as well. For $(S_L,\,S_R)$ type NP couplings, the allowed ranges
in $R_D^{\tau}$, $R_{D_s}^{\tau}$, and $R_{\pi}^{\tau}$ reduce considerably, whereas, 
there is no or very little change in $R_{D^{\ast}}^{\tau}$, $R_{D^{\ast}_s}^{\tau}$, $R_{D_s}$,
$R_{D^{\ast}_s}$, and $R_{\pi}$ allowed ranges once additional $2\sigma$
constraints from the estimated values of $R_D^{\tau}$ and $R_{D^{\ast}}^{\tau}$ are imposed. Similar results are obtained
for $(\widetilde{S}_L,\,\widetilde{S}_R)$ NP couplings as well.

Although, current measurements from BABAR and BELLE suggest presence of NP, NP is yet to be confirmed. Both experimental and 
theoretical precision in these $B$ decay modes are necessary for a reliable interpretation of NP signals if NP is indeed 
present.
Retaining our current
approach, we could sharpen our estimates once improved measurement of $V_{ub}$ is available. These newly defined observables 
may, in future, play a crucial role in identifying the nature of NP couplings in $b \to (u,\,c)\tau\nu$ decays.
Again, precise data on $\mathcal B(B \to \pi\,\tau\,\nu)$ will put additional constraint on the NP parameter space. 
Similarly, measurement of 
$R_{D_s}$ and $R_{D^{\ast}_s}$ will also help in identifying the nature of NP couplings in $b \to c\,\tau\nu$ decays.

\end{document}